\documentclass[superscriptaddress,showpacs,amssymb,10pt,reprint,aps,prd,longbibliography,nofootinbib,floatfix]{revtex4-2}

\usepackage{graphicx,epsfig,amssymb,times} 
\usepackage{amsmath,amsfonts}
\usepackage{bm}
\usepackage{epstopdf}
\usepackage{hyperref}
\usepackage{float}
\usepackage[usenames]{color}   
\usepackage[dvipsnames]{xcolor}
\usepackage[normalem]{ulem}
\usepackage[titletoc]{appendix}
\usepackage{subfigure}
 
\definecolor{coolblack}{rgb}{0.0, 0.18, 0.39}
\definecolor{darkred}{rgb}{0.5,0,0}
\definecolor{darkgreen}{rgb}{0,0.5,0}
\definecolor{darkblue}{rgb}{0,0,0.5}
\definecolor{lapislazuli}{rgb}{0.15, 0.38, 0.61}
\definecolor{venetianred}{rgb}{0.78, 0.03, 0.08}
\definecolor{bleudefrance}{rgb}{0.19, 0.55, 0.91}
\definecolor{dogwoodrose}{rgb}{0.84, 0.09, 0.41}
\hypersetup{colorlinks=true, citecolor=darkblue, linkcolor=darkblue, 
urlcolor = darkblue}

\begin{document}

\title{\large Absorption and scattering spectra of massive scalar waves in charged regular black hole spacetimes}
	
	\author{Marco A. A. de Paula}
	\email{marcodepaula@ufpa.br}
	\affiliation{Faculdade de Ciências Naturais, Universidade Federal do Par\'a, Campus Universit\'ario do Tocantins-Cametá, 68400-000, Camet\'a, Par\'a, Brazil}
	\affiliation{Programa de P\'os-Gradua\c{c}\~{a}o em F\'{\i}sica, Universidade 
		Federal do Par\'a, 66075-110, Bel\'em, Par\'a, Brazil.}

	\author{Carolina L. Benone}
	\email{benone@ufpa.br}
	\affiliation{Campus Salin\'opolis, Universidade Federal do Par\'a, 68721-000, Salin\'opolis, Pará, Brazil.}
	
	\author{Lu\'is C. B. Crispino}
	\email{crispino@ufpa.br}
	\affiliation{Programa de P\'os-Gradua\c{c}\~{a}o em F\'{\i}sica, Universidade 
		Federal do Par\'a, 66075-110, Bel\'em, Par\'a, Brazil.}

\begin{abstract}

Regular black holes (RBHs) can be seen as possible alternatives to standard black holes (BHs), since these geometries do not have a curvature singularity. As a way of improving our knowledge of such geometries, we can investigate how the astrophysical environment interacts with RBHs and compare the results with those obtained in the framework of standard BHs. We aim to study the absorption and scattering cross sections of massive scalar waves impinging on Ayón-Beato-García and Bardeen charged RBH geometries, focusing on understanding the role played by the field's mass. Concerning the absorption spectrum, our numerical results show that the total absorption cross section increases as we increase the field's mass for fixed values of the BH charge. In turn, in the scattering spectrum, an increase in the mass of the field leads to wider interference widths for field velocities larger than a critical value, $v_c$, and the scattering intensity flux increases as we consider higher field mass values. Moreover, we compare our numerical results with the classical and semiclassical approximations, showing that they agree very well within the appropriate limits. We also draw comparisons with the results of the Reissner-Nordström metric. In particular, we show that the mass of the field contributes to finding situations in which the absorption and scattering spectra of regular and standard BHs are similar for arbitrary values of the field frequency and scattering angle, considering low- to near-extreme BH charges. 

\end{abstract}

\date{\today}

\maketitle

\section{Introduction}

The concept of spacetime singularities has puzzled the scientific community since the early days of standard general relativity (GR). Throughout the years, several definitions and formulations have appeared in the literature; see, e.g., Refs.~\cite{HE1973,S2006,TCE1980} for an enlightening discussion on the concept of spacetime singularities and some definitions. One can split singularities into two distinct classes: curvature and noncurvature singularities~\cite{W1984}. There is also a softer formulation in which singularities are divided into curvature singularities and coordinate singularities~\cite{IV2022}. In the former, the known laws of physics break down since physical and geometrical quantities diverge. Meanwhile, coordinate singularities, such as the event horizon radius in the Schwarzschild coordinates, are pathologies of the coordinate systems that can be circumvented by a suitable coordinate transformation. For example, by using the Painlevé-Gullstrand coordinates we can obtain a well-defined spacetime metric at the event horizon~\cite{P1921,G1922,MP2001}.  

Standard GR admits the existence of black hole (BH)  spacetimes without curvature singularities. In this respect, a good strategy is to couple different matter sources with gravity. These couplings can lead to the so-called regular black hole (RBH) solutions, which are curvature singularity-free BH geometries. The first model of RBH was proposed as a toy model by James Bardeen (BD) in 1968~\cite{B1968}. Thirty years later, Eloy Ay\'on-Beato and Alberto Garc\'ia (ABG) found that nonlinear electrodynamics (NED) models minimally coupled to GR can be used to obtain electrically charged RBH solutions~\cite{ABG1998}. The NED theory is a possible well-motivated generalization of Maxwell's linear electrodynamics in the strong electromagnetic field regime~\cite{B1934,BI1934,HGP1987,FT1985,SW1999,A2000,NBS2004,RPM2017,ATLAS2017,ATLAS2019,PVLAS2020}.

In the so-called $F$ framework~\cite{B2001}, the action that governs the NED theory minimally coupled to GR can be written as
\begin{equation}
\label{S}\mathcal{S} = \dfrac{1}{16\pi}\int \sqrt{-g} \left[ R-\mathcal{L}(F) \right]d^{4}x, 
\end{equation}
where $R$ is the Ricci scalar, $\mathcal{L}(F)$ is a gauge-invariant electromagnetic Lagrangian density, with $F$ being the Maxwell scalar, and $g$ is the determinant of the metric tensor $g_{\mu \nu}$. An alternative form for action~\eqref{S}, known as the $P$ \textit{framework}~\cite{B2001}, can be obtained by introducing a structural function, namely
\begin{equation}
\mathcal{H}(P) \equiv 2F\mathcal{L}_{F} - \mathcal{L}(F),
\end{equation}
via Legendre transformation~\cite{HGP1987}, where $P$ is an electromagnetic scalar defined in this framework. We can find exact charged RBH solutions by providing the NED model and solving the corresponding field equations. Nowadays, there are several charged RBH solutions based on the minimal coupling between GR and NED sources (see, e.g., Refs.~\cite{B2001,D2004,BV2014,K2017,ABG2000,KAB2023} and references therein), including the geometries of Bardeen~\cite{ABG2000,RS2018} and Hayward~\cite{FW2016,TSA2018}. There are also NED models related to hairy BHs obtained via gravitational decoupling~\cite{JO2021}, and solutions beyond GR~\cite{OR2011,JRH2015,SR2018}.

In real astrophysical scenarios, the interaction of the BH with its environment is a crucial issue, since BHs are expected to be surrounded by distributions of matter~\cite{N2005}. A standard procedure to improve our understanding of BH physics is to study how they absorb or scatter matter fields. The pioneering studies on this subject were performed in the 1970s, considering BHs in vacuum (cf., e.g., Ref.~\cite{FHM1988,NS1977,NS1978,NS1978b} and references therein). During the past fifty years, the absorption and scattering cross sections of standard BH solutions were computed in many distinct situations (cf., e.g., Refs.~\cite{JP2004,SRD2008,CDE2009,OCH2011,CB2014,CBM2013,LBC2017,BC2019,OCH2011,FH2020,HSD2020,LWJ2025} and references therein). Recently, a few studies on how static and spherically symmetric RBH solutions 
 absorb and scatter test matter fields have been done~\cite{MC2014,MOC2015,S2017,SBP2018,PLC2020,PLC2022,PLC2023,BP2025,THZ2026,RK2025,PLC2023b,TPLA2026}. 
However, these works mainly consider massless test scalar fields. Therefore, the role of the mass in the absorption and scattering spectra of massive test scalar fields in the background of charged RBH spacetimes has not yet been extensively investigated.

We study the propagation of massive test scalar fields in the background of ABG and BD RBH solutions, which are static and spherically symmetric charged RBH geometries based on NED sources. We focus on the absorption and scattering patterns aiming to understand the contributions of the field mass to the absorption and scattering properties of the RBHs. We also address the possibility of RBHs mimicking the absorption and scattering properties of standard BHs, such as the Reissner-Nordstr\"om (RN) one. Moreover, we point out that the absorption cross section (ACS) of massive test scalar fields in the RN background has already been analyzed in Ref.~\cite{CB2014}, and the differential scattering cross section (SCS) was partially studied in Ref.~\cite{SL2021}. Therefore, for completeness, here we also revisit the absorption and scattering spectra of massive test scalar fields in the RN spacetime in detail.

The remainder of this work is organized as follows. In Sec.~\ref{sec:bg}, we briefly review ABG, BD, and RN BH spacetimes. In Sec.~\ref{sec:ga}, we study the geodesic equations, presenting analytical approximations for the absorption and scattering cross sections in the high-frequency regime. In this section, we also extend the analysis of the glory approximation made for bosonic fields in Schwarzschild spacetime~\cite{PDC2024b} to charged geometries. The propagation of massive test scalar fields in the background of the BH geometries is investigated in Sec.~\ref{sec:sf}, where we also introduce the main equations to compute the absorption and scattering cross sections. In Sec.~\ref{sec:rd}, we exhibit a selection of our main results concerning the absorption and scattering patterns of ABG, BD, and RN BH spacetimes. We then present our concluding remarks in Sec.~\ref{sec:remarks}. Throughout this paper, we use natural units, for which $G = c = \hbar = 1$, and the metric signature $+2$ $(- + + +)$.

\section{Background}\label{sec:bg}

The spacetime configuration is static and spherically symmetric, with line element
\begin{equation}
\label{SC}ds^{2} = -f(r)dt^{2}+f(r)^{-1}dr^{2}+r^{2}\left(d\theta^{2}+\sin^{2}\theta d\varphi^{2}\right),
\end{equation}
where $f(r)$ is the metric function. The corresponding metric functions of ABG~\cite{ABG1998}, BD~\cite{ABG2000} and RN~\cite{W1984} spacetimes are
\begin{subequations}
\begin{align}
\label{MF_ABG}f^{\rm{ABG}}(r) = \ & 1 - \dfrac{2Mr^{2}}{(r^{2}+Q^{2})^{3/2}}+\dfrac{Q^{2}r^{2}}{(r^{2}+Q^{2})^{2}}, \\
\label{MF_BD}f^{\rm{BD}}(r) =  \ & 1 - \dfrac{2Mr^{2}}{(r^{2}+Q^{2})^{3/2}}, \\
\label{MF_RN}f^{\rm{RN}}(r) =  \ & 1 - \dfrac{2M}{r} + \dfrac{Q^{2}}{r^{2}},
\end{align}
\end{subequations}
respectively. The parameters $M$ and $Q$ correspond to the mass and charge of the central object, respectively. Throughout this work, we use the same symbol ($Q$) for both magnetic (BD case) and electric (ABG and RN cases) charges.

The ABG solution behaves asymptotically as the RN solution, unlike the BD solution, which behaves as
\begin{equation}
\label{LC_BD}f^{\rm{BD}}(r \rightarrow \infty) = 1 - \dfrac{2M}{r}+\dfrac{3MQ^{2}}{r^{3}}+\mathcal{O}\left[\dfrac{1}{r^{5}}\right].
\end{equation}
In the metric functions~\eqref{MF_ABG}-\eqref{MF_RN}, the limit $Q \rightarrow 0$ leads to the Schwarzschild metric function, namely
\begin{equation}
\label{MF_Sch}f^{\rm{SCH}}(r) = 1 - \dfrac{2M}{r}.
\end{equation}

By solving $f(r) = 0$ we can determine the location of the horizons. The extreme charge value, denoted by $Q_{\rm{ext}}$, can be obtained by solving $f(r) = 0$ and $f^{\prime}(r) = 0$ simultaneously, where the prime symbol $^\prime$ denotes differentiation with respect to the radial coordinate $r$. Therefore, we can show that the values of $Q_{\rm{ext}}$ for ABG, BD, and RN solutions are given by
\begin{equation}
Q_{\rm{ext}}^{\rm{ABG}} \cong 0.6341 M, \ \ Q_{\rm{ext}}^{\rm{BD}} = \dfrac{4M}{3\sqrt{3}}, \ \ \text{and} \ \ Q_{\rm{ext}}^{\rm{RN}} = M, 
\end{equation}
respectively.

\begin{figure}[!htbp]
\begin{centering}
    \includegraphics[width=1.0\columnwidth]{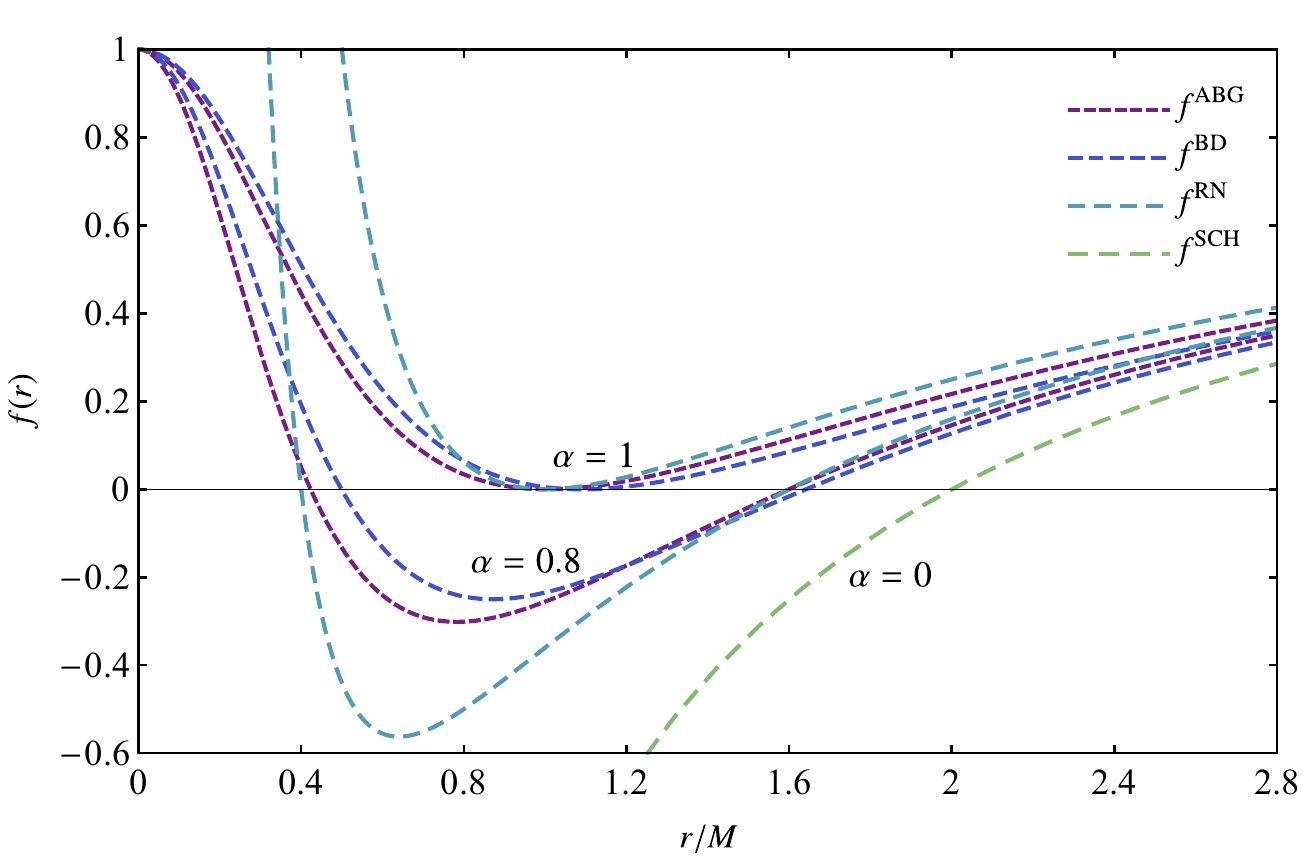}
    \caption{Comparison between the metric functions of ABG, BD, and RN BH geometries for two distinct values of $\alpha$, namely $0.8$ and $1$. We also exhibit the Schwarzschild case ($\alpha = 0$) for comparison.}
    \label{mf}
\end{centering}
\end{figure}

To facilitate comparisons between different spacetimes, it is useful to introduce the normalized charge, defined as
\begin{equation}
\alpha \equiv \dfrac{Q}{Q_{\rm{ext}}},
\end{equation}
which satisfies $0 \leq \alpha \leq 1$ for BH geometries. As shown in Fig.~\ref{mf}, the ABG, BD, and RN spacetimes have a similar causal structure. For $\alpha < 1$, we have the Cauchy $r_{-}$ and event horizons $r_{+}$, while for $\alpha = 1$, we have an extremely charged BH, for which $r_{+} = r_{-}$. In turn, $\alpha > 1$ is related to horizonless solutions. The case $\alpha = 0$ corresponds to the Schwarzschild spacetime. Here, we consider only BH geometries.

\section{Geodesic Analysis}\label{sec:ga}

In this section, we investigate the motion of uncharged massive particles in the setup introduced in Sec.~\ref{sec:bg}. We also present classical and semiclassical analytical approximations for the absorption and differential scattering cross sections.

\subsection{Motion of timelike particles}

We can show that uncharged massive particles with mass $m$ satisfy the following radial equation~\cite{C2019}
\begin{equation}
\label{RE}\left(\dfrac{m^{2}}{L^{2}}\right)\dot{r}^{2} = \dfrac{E^{2}}{L^{2}}-f(r)\left(\dfrac{m^{2}}{L^{2}}+\dfrac{1}{r^{2}}\right),
\end{equation}
where the overdot stands for the derivative with respect to the proper time, and we considered, without loss of generalization, the motion in the equatorial plane, i.e., $\theta = \pi /2$. The quantities $E$ and $L$ are the energy and angular momentum of the particle, respectively, which at the semiclassical limit can be mapped as $E \rightarrow \omega$ and $L \rightarrow l + 1/2$, respectively.

By defining the impact parameter $b\equiv L/v E$ and $\mathcal{K}(r) \equiv \dot{r}^{2}(m^2/L^2)$, we can rewrite Eq. \eqref{RE} as 
\begin{equation}
\label{RE1}\mathcal{K}(r) = \dfrac{1}{{b^2v^2}}-f(r)\left(\frac{1-v^2}{b^2v^2}+\dfrac{1}{r^{2}}\right),
\end{equation}
where $v$ is the particle speed (in units of $c$), namely,~\cite{SD2007}
\begin{equation}
\label{ADP}v \equiv\sqrt{1-\frac{m^{2}}{E^2}}.
\end{equation}

An unstable circular orbit with radius $r_{c}$ satisfies the following pair of equations: $\mathcal{K}(r_{c}) = 0$ and $\mathcal{K}^{\prime}(r_{c}) = 0$. By solving them, we may find the critical impact parameter $b_{c} \equiv L_{c}/vE_{c}$, given by 
\begin{equation}
\label{CIP}b_{c} = \dfrac{r_{c}}{v f_{c}^{1/2}}\sqrt{1-\left(1-v^2\right) f_{c}},
\end{equation}
where $f_{c} \equiv f(r_{c})$, and an equation that provides the values of $r_{c}$, which is given by
\begin{equation}
\label{eqlr}\dfrac{df(r)}{dr}\bigg|_{r=r_{c}} = \dfrac{2f_{c}\left(1-(1-v^{2})f_{c}\right)}{r_{c}}.
\end{equation}
Notice that for $m = 0$, we have $v = 1$, and Eqs.~\eqref{CIP} and~\eqref{eqlr} reduce to their massless counterparts.

By evaluating Eq.~\eqref{eqlr} numerically, we can find the values of $r_{c}$ and consequently $b_{c}$. The classical capture cross section of timelike geodesics, also known as the geometric cross section (GCS), is then given by~\cite{W1984}
\begin{equation}
\label{GCS}\sigma_{\rm{gcs}} = \pi b_{c}^{2}.
\end{equation}

The deflection angle of the scattered massive particle can be written as~\cite{N2013}
\begin{equation}
\label{DA}\Theta(b) = 2\int_{r_{0}}^{\infty} \dfrac{1}{\mathcal{U}(r)}dr - \pi,
\end{equation} 
where
\begin{equation}
\label{U(r)}\mathcal{U}(r) \equiv \dfrac{dr}{d \varphi} = r^{2}\sqrt{\dfrac{1}{{b^2v^2}}-f(r)\left(\frac{1-v^2}{b^2v^2}+\dfrac{1}{r^{2}}\right)},
\end{equation}
and $r_0$ is the turning point of the particle, which satisfies
\begin{equation}
\mathcal{U}(r)|_{r = r_{0}} = 0.
\end{equation}

We can obtain an analytic expression for the deflection angle in the weak field limit by using the geodesic method. We expand the integrand of Eq.~\eqref{DA} in powers of $1/r$. The radius $r_{0}$ as a function of $b$ is obtained by solving Eq.~\eqref{U(r)} and expanding the results in powers of $2M/b$. Following these steps, we obtain that the weak deflection angle of massive particles in the background of ABG, BD, and RN spacetimes is given by 
\begin{align}
\label{thetaABG} \Theta(b)_{i} =  \dfrac{2Mz}{b} + \mathcal{O}\left[\dfrac{1}{b^{2}}\right],
\end{align}
where $i = (\text{ABG}, \text{BD}, \text{RN})$,  and we considered only the contributions up to the first order of $1/b$, which is the relevant order for our purposes. We also defined $z \equiv z(v)$ as
\begin{equation}
z(v) \equiv  1+\dfrac{1}{v^2}.
\end{equation}

From Eq.~\eqref{thetaABG}, we see that the contributions from the mass of the particle already appear in the first order, and the weak deflection angle diverges in the limit $m \rightarrow E$. We also note that the weak deflection angle of the ABG, BD, and RN solutions coincides in the first order, as expected, since the metric function of the ABG and BD geometries tends to $1-2M/r + \mathcal{O}[1/r^{2}]$ in the far field. Moreover, the charge contributions are negligible far away from the BH, since they do not modify the dominant term. In particular, we can show that the charge contributions in the BD case appear only for orders higher than $1/b^{2}$ due to the asymptotic behavior of the BD geometry [cf. Eq.~\eqref{LC_BD}]. We also note that in the massless limit $z(v) \rightarrow 2$, and we recover the Einstein's deflection angle, namely $\Theta(b) = 4M/b$~\cite{W1984}.

The classical differential SCS is given by~\cite{N2013}
\begin{equation}
\label{CSCS}\dfrac{d\sigma_{\rm{cl}}}{d\Omega} = \dfrac{1}{\sin\theta}\sum_{n}b(\theta)\bigg|\dfrac{db(\theta)}{d\theta}\bigg|,
\end{equation}
where $\theta$ is the scattering angle, which is related to the deflection angle by $\theta=\Theta-2n\pi$, with $n\in\mathbb{Z}^{+}$. The parameter $n$ measures the number of times that the massive particle rotates around the BH before being scattered to infinity. The classical differential SCS can be obtained by inverting Eq.~\eqref{DA} and inserting $b(\theta)$ into Eq.~\eqref{CSCS}. In the weak field limit, we can use Eqs.~\eqref{thetaABG} to obtain the differential SCS for small angles, which can be expressed as
\begin{align}
\label{CSCSweakABG} \dfrac{d\sigma_{\rm{cl}}^{i}}{d\Omega} = \dfrac{4M^2z^{2}}{\Theta^4} + \mathcal{O}\left[\dfrac{1}{\Theta^{3}}\right],
\end{align}
where we have considered only the main contribution to $1/\Theta$. The classical differential SCS is independent of the BH charge in the high-frequency limit. We also notice that the mass of the particle plays an important role in this limit, since the classical differential SCS diverges as $m \rightarrow E$ (or, equivalently, as $v \rightarrow 0$). Furthermore, the classical differential SCS of the ABG and BD RBHs tends to that of the RN BH in the leading order of the differential SCS, as expected. The results for the massless case are properly obtained in the limit $m \rightarrow 0$ (or, equivalently, as $v \rightarrow 1$).

\begin{figure}[!htpb]
\begin{centering}
    \includegraphics[width=1.0\columnwidth]{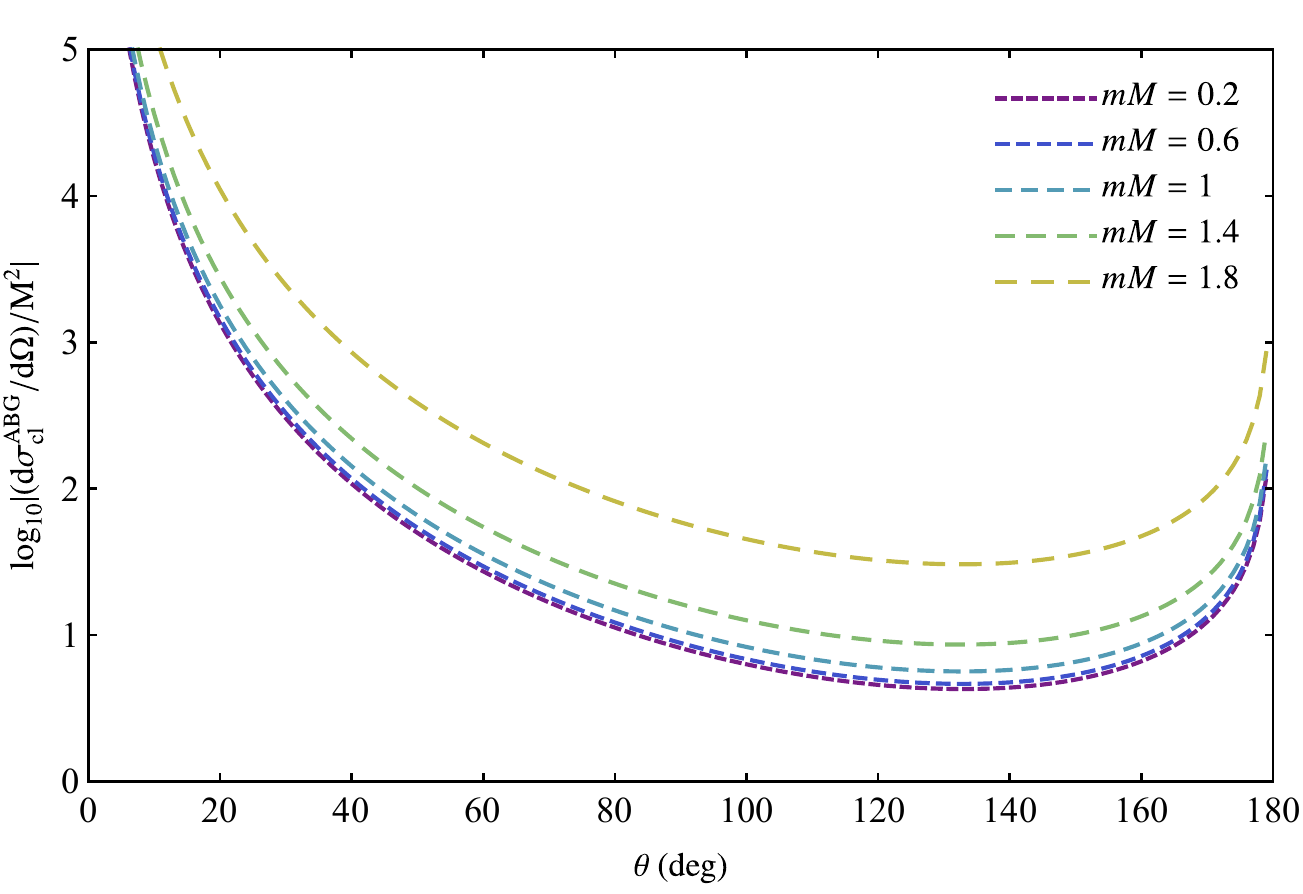}
    \caption{Classical differential SCS of the ABG RBH, considering different values of $mM$ for $\alpha = 0.8$ and $EM = 2$.}
    \label{CSC}
\end{centering}
\end{figure}

In Fig.~\ref{CSC}, we exhibit the classical differential SCS of the ABG RBH. We note that the classical differential SCS increases as we consider higher values of $mM$ and the contributions of the mass intensify as $m M \rightarrow E M$, which can be seen from the enlargement in the spacing between adjacent curves. We also point out that the results for the BD and RN BHs are qualitatively similar.

\begin{figure}[!htpb]
\begin{centering}
    \includegraphics[width=1.0\columnwidth]{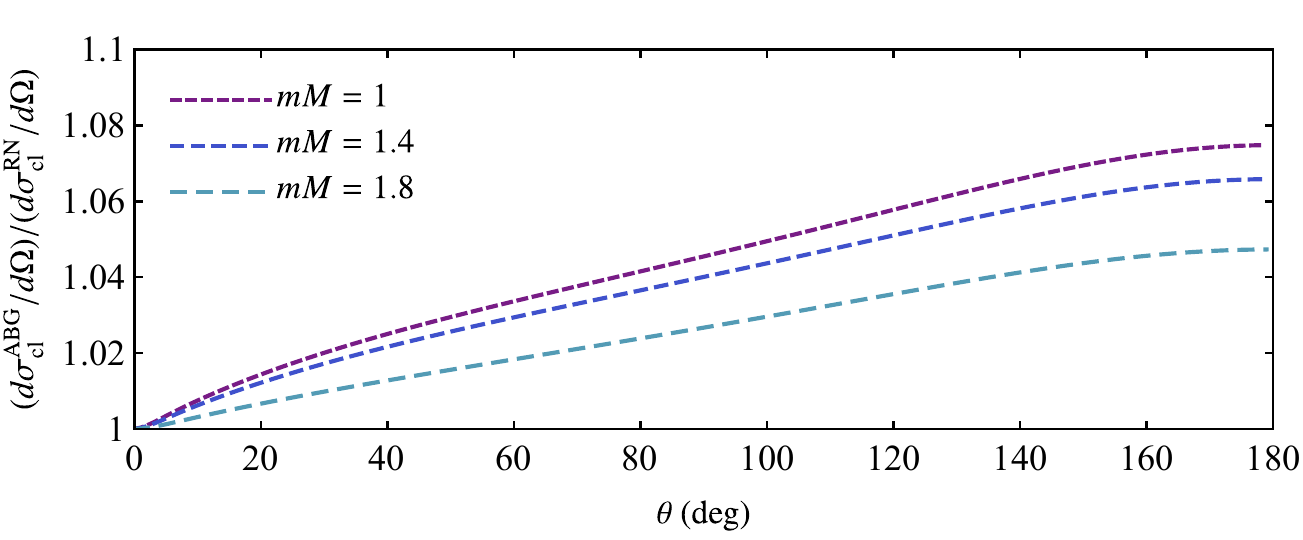}
    \includegraphics[width=1.0\columnwidth]{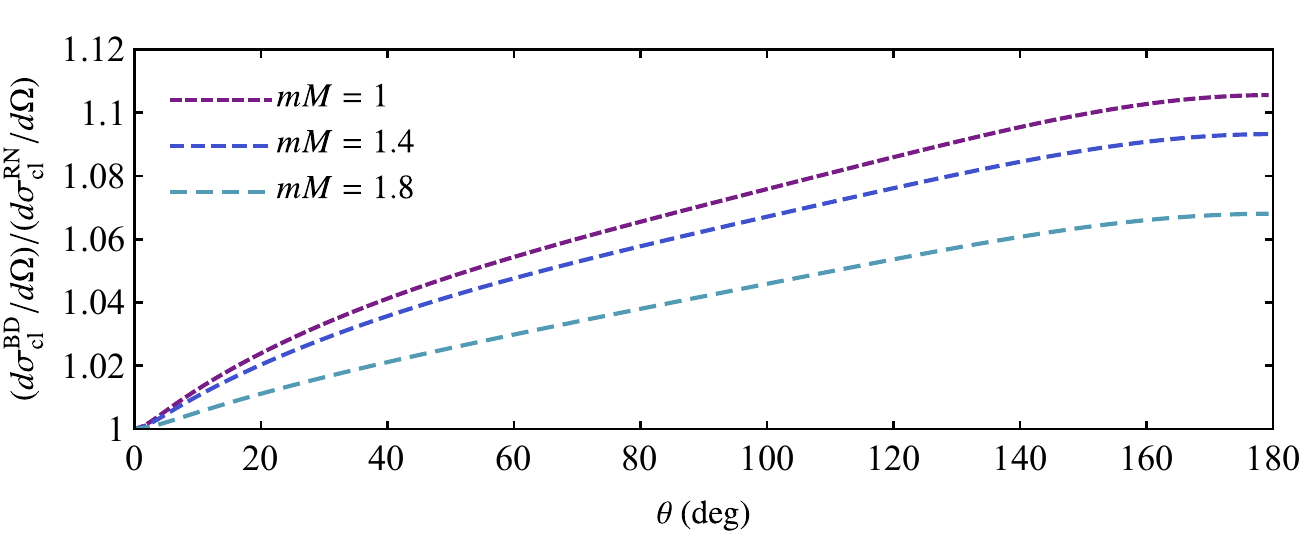}
    \includegraphics[width=1.0\columnwidth]{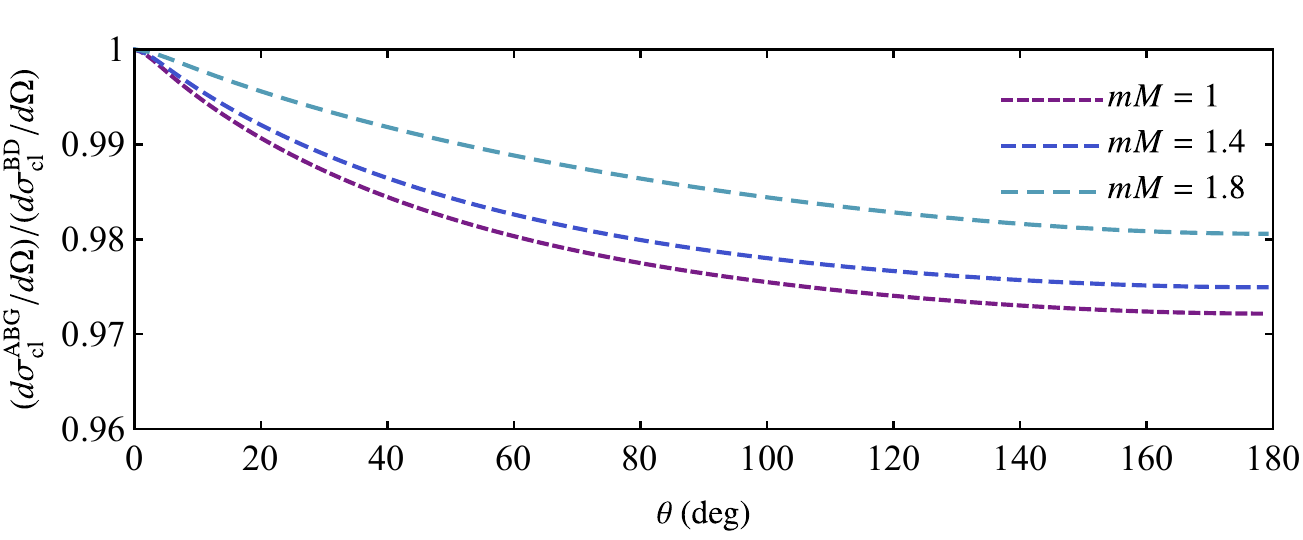}
    \caption{Ratio between the classical differential SCSs of ABG, BD, and RN BHs, considering distinct values of $mM$, for $\alpha = 0.8$ and $EM = 2$. From the top to the bottom panel, we consider the ratio between ABG and RN BHs, BD and RN BHs, and ABG and BD RBHs, respectively.}
    \label{CSC2}
\end{centering}
\end{figure}

In Fig.~\ref{CSC2}, we compare the classical differential SCSs of ABG, BD, and RN BHs. As we can see, for fixed values of $\alpha$ and $mM$, we find that
\begin{equation}
\label{CSCcomp}\dfrac{d\sigma_{\rm{cl}}^{\rm{BD}}}{d\Omega} > \dfrac{d\sigma^{\rm{ABG}}_{\rm{cl}}}{d\Omega} > \dfrac{d\sigma_{\rm{cl}}^{\rm{RN}}}{d\Omega}.
\end{equation}

\subsection{Semiclassical approximations}

In the high-frequency regime, we can address the motion of massive scalar waves as that of massive particles, provided that, in this limit, the wavelength of the scalar field is very small when compared to the scale of the BH. Therefore, in this frequency regime, the scalar wave follows timelike geodesics that satisfy Eq.~\eqref{RE}, and the GCS is given by Eq.~\eqref{GCS}. Besides that, in the absorption scenario, we can use a semiclassical approximation, known as sinc approximation, to capture wave interference effects related to the interaction of the scalar wave.

For massive scalar waves with mass $\mu$ and frequency $\omega$, this approximation can be written as~\cite{CB2014}
\begin{equation}
\label{SINC}\sigma_{\rm{hf}} \approx \sigma_{\rm{gcs}}\left[1-8\pi v b_{c} \Lambda e^{-\pi v b_{c} \Lambda} \text{sinc}(2\pi v b_{c} \omega)\right],
\end{equation}
where $\text{sinc}(x) = \sin(x)/x$ and $\Lambda$ is the Lyapunov exponent related to the circular orbit~\cite{VC2009}, given by
\begin{equation}
\label{lyapu}\Lambda = f_{c}b_{c}v\sqrt{\dfrac{\mathcal{K}^{\prime\prime}_{c}}{2}},
\end{equation}
for timelike geodesics, in which $\mathcal{K}^{\prime\prime}_{c} \equiv \mathcal{K}^{\prime\prime}(r_{c})$ is given by
\begin{equation}
\label{Vr}\mathcal{K}^{\prime\prime}_{c}=\dfrac{2f_{c}\left(1-4(1-v^{2})f_{c}\right)}{r_{c}^{4}}-\dfrac{f^{\prime\prime}_{c}}{r_{c}^{2}\left(1-(1-v^{2})f_c \right)}.
\end{equation}
Recall that, in this context, the parameter $v$ is interpreted as the ratio of the speed of propagation of a scalar wave in the far-field to the speed of light with $m \rightarrow \mu$ and $E \rightarrow \omega$.

\begin{figure}[!htbp]
\begin{centering}
    \includegraphics[width=1.0\columnwidth]{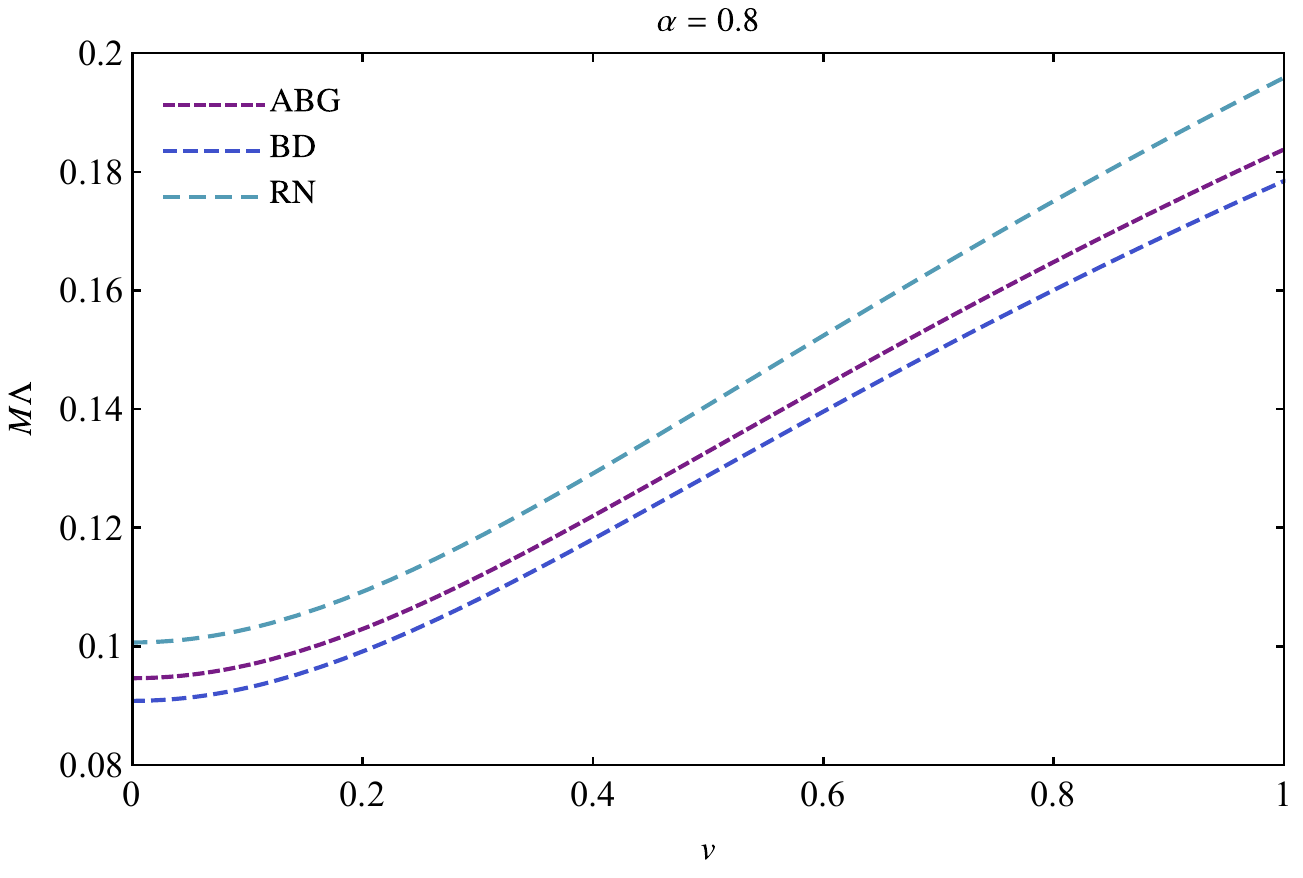}
	\includegraphics[width=1.0\columnwidth]{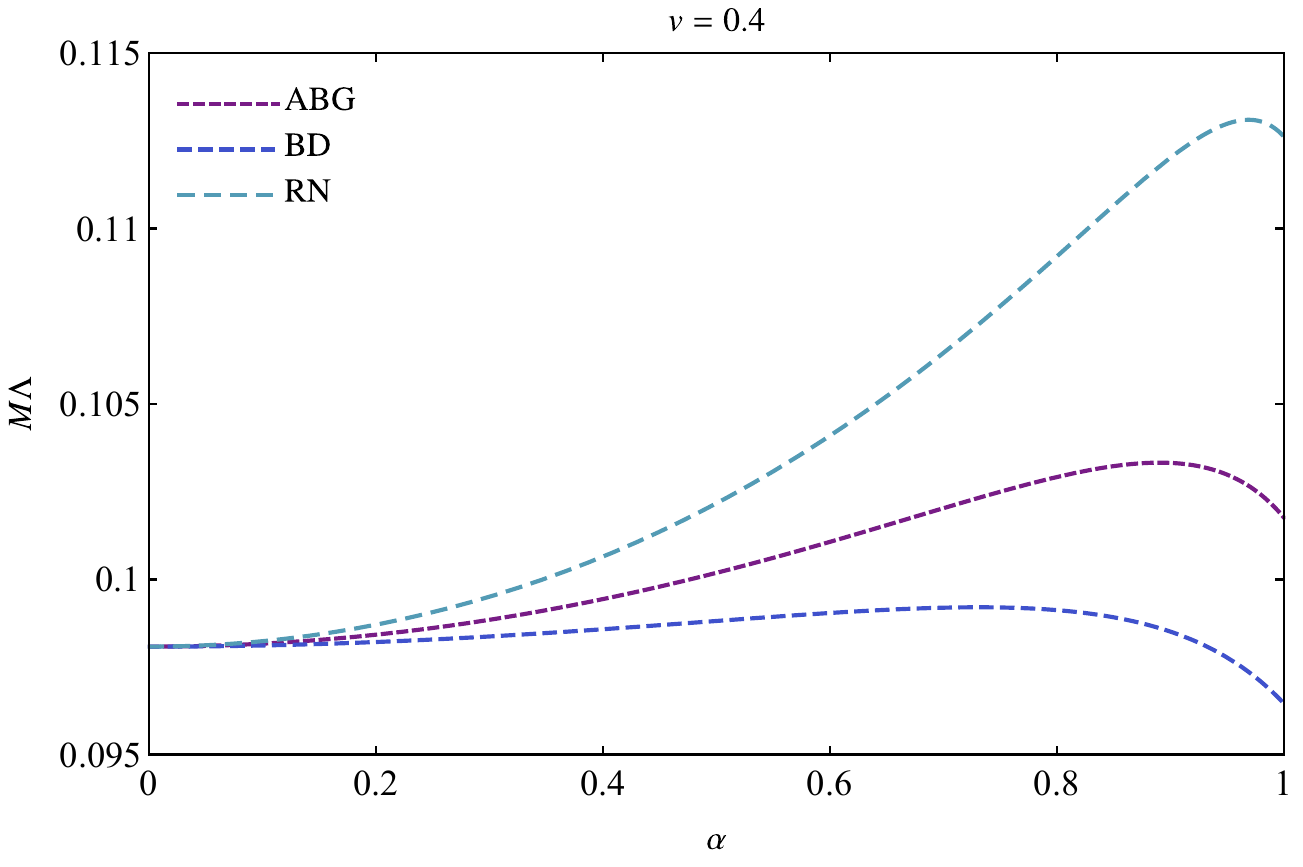}
    \caption{Lyapunov exponent for ABG, BD, and RN BHs, considering two distinct scenarios: (i) different values of $v$ for $\alpha = 0.8$ (top panel); and (ii) distinct choices of $\alpha$ for $v = 0.4$ (bottom panel).}
    \label{lyapunov}
\end{centering}
\end{figure}

For comparison purposes, we exhibit the behavior of the Lyapunov exponent in Fig.~\ref{lyapunov}. We notice that the pattern of the Lyapunov exponent for the ABG, BD, and RN BH geometries is similar, but its magnitude is different. As we can see, the Lyapunov exponent monotonically increases as we consider higher values of $v$ for fixed values of $\alpha$. However, for sufficiently slow massive scalar fields, we observe that the Lyapunov exponent does not increase monotonically as $\alpha$ increases. Notice that the instability timescale of the orbit is associated with the Lyapunov exponent (see, e.g., Refs.~\cite{VC2009,BC1991,CL2003}), i.e., $\tau \equiv 1/\Lambda$. Therefore, we conclude that increasing (decreasing) values of $\Lambda$ are related to an instability timescale that gets shorter (longer). In particular, we point out that the scenario characterized by the bottom panel of Fig.~\ref{lyapunov} is absent for RBHs when one considers massless scalar waves (see, e.g., Fig. 8 of Ref.~\cite{PLC2020}). Therefore, the mass contribution of the scalar field affects the instability timescale with a very distinctive signature and no counterpart for RBHs when considering massless scalar fields. Moreover, the results obtained in the massless case are recovered in the limit $v \rightarrow 1$. 

For the scattering case, the classical differential SCS is given by Eq.~\eqref{CSCS}. That approximation is not useful for forecasting the interference effects associated with the scalar wave propagation. To reveal some of these properties, we can use the semiclassical glory approximation, which typically works very well near the backward direction $(\theta = \pi)$. In the background of static and spherically symmetric BH geometries, the glory approximation can be written as~\cite{RAM1985,SD2007}
\begin{equation}
\label{glory}\dfrac{d\sigma_{\rm{g}}}{d\Omega} = 2\pi \omega v b_{g}^{2}\bigg|\dfrac{db}{d\theta}\bigg|_{\theta = \pi}J_{0}^{2}(\omega v b_{g}\sin\theta),
\end{equation}
where $b_{g}$ is the impact parameter of the backscattered rays and $J_{0}$ is the Bessel function of the first kind.

\begin{figure}[!htbp]
  \centering
   \includegraphics[width=1\columnwidth]{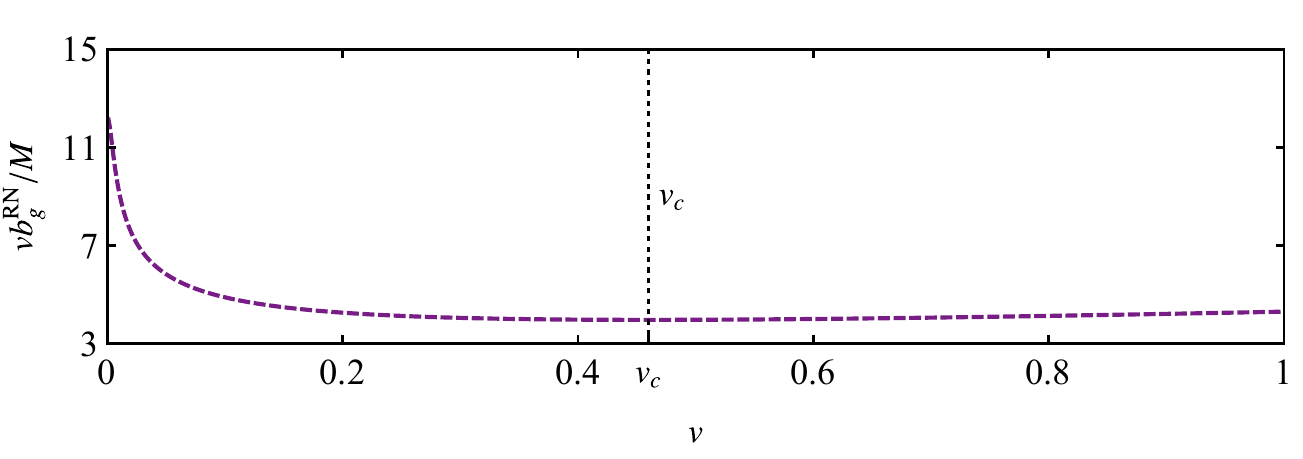}\label{vbgrn}
\caption{Behavior of the coupling $vb_{g}$ for the RN BH, as a function of $v$, considering $\alpha = 1$. In this case, we have $v_{c}^{\rm{RN}} \approx 0.459$.}
\label{vbg}
\end{figure}

We notice that the interference fringe width is inversely proportional to the argument of the Bessel function $(\sim 1/\omega vb_{g})$, and the backscattered flux intensity is proportional to $b_ {g}^{2}|db/d\theta|_{\theta = \pi}$. Moreover, in Ref.~\cite{PDC2024b}, it is shown for the Schwarzschild case that $vb_{g}$ is maximal in the speedless limit, decreases as we increase $v$ until a critical velocity value $v_{c}$, and then starts to increase. The critical velocity corresponds to the situations where
\begin{equation}
\dfrac{d\left(vb_{g}\right)}{dv}\bigg|_{v = v_{c}} = 0.
\end{equation}
In Fig.~\ref{vbg}, we exhibit the behavior of $vb_{g}$ for the RN BH solution. Our results indicate that it presents a pattern similar to that of the Schwarzschild case, regardless of the value of the BH charge. Therefore, we expect that, for fixed frequency values, the interference fringe width gets wider (narrower) as we consider heavier fields for $v > v_{c}$ $(v < v_{c})$. We recall that an increase in the field velocity is equivalent to a decrease in the field mass, for fixed frequency values. Moreover, the results for the RBHs are qualitatively similar. The main difference is the location of $v_c$, which is slightly different for each BH solution. For example, for extremely charged ABG and BD RBHs, we have $v_{c}^{\rm{ABG}} \approx 0.441$ and $v_{c}^{\rm{BD}} \approx 0.422$, respectively, in contrast to $v_{c}^{\rm{RN}} \approx 0.459$, for the RN case.

In Sec.~\ref{sec:rd}, we compare the classical [cf. Eqs.~\eqref{GCS} and~\eqref{CSCS}] and semiclassical approximations [cf. Eqs.~\eqref{SINC} and~\eqref{glory}] with our numerical results.

\section{Scalar field}\label{sec:sf}

In this section, we study the dynamics of massive scalar waves in the background presented in Sec.~\ref{sec:bg}. In Sec.~\ref{subsec:msw}, we analyze the main equations that govern the dynamics and the corresponding effective potential. In Sec.~\ref{subsec:abs}, we present the expressions that will be used to compute the absorption and differential scattering cross sections.

\subsection{Massive scalar waves}\label{subsec:msw}

The Klein-Gordon equation that governs the propagation of an uncharged massive scalar field $\Phi$, with mass $\mu$, in the background of curved spacetimes is given by
\begin{equation}
\label{KG}\dfrac{1}{\sqrt{-g}}\partial_{\mu}\left(\sqrt{-g}g^{\mu \nu}\partial_{\nu}\Phi\right)-\mu^{2}\Phi = 0.
\end{equation}
The field $\Phi$ can be decomposed as
\begin{equation}
\label{PHI}\Phi= \dfrac{1}{r}\sum_{l = 0}^{\infty} C_{\omega l}\Psi_{\omega l}(r)P_{l}(\cos\theta)e^{-i\omega t},
\end{equation}
where $C_{\omega l}$ are constant coefficients, with $\omega$ and $l$ being the frequency and angular momentum of the partial waves, respectively. The functions $P_{l}$ are the Legendre polynomials and $\Psi_{\omega l}$ are the radial functions. Using the tortoise coordinate $r_{\star}$, defined as $f(r)dr_{\star}=dr$, we can show that $\Psi_{\omega l}$ satisfies 
\begin{equation}
\label{RE_TC}\frac{d^{2}}{dr_{\star}^{2}}\Psi_{\omega l}+\left[\omega^{2}-V_{\rm{eff}}(r)\right]\Psi_{\omega l}=0,
\end{equation}
where the effective potential $V_{\rm{eff}}(r)$ reads
\begin{equation}
\label{EP}V_{\rm{eff}}(r) = f(r)\left(\mu^{2}+\dfrac{f^{\prime}(r)}{r}+\dfrac{l(l+1)		}{r^{2}}\right).
\end{equation}

\begin{figure}[!htbp]
\begin{centering}
    \includegraphics[width=1.0\columnwidth]{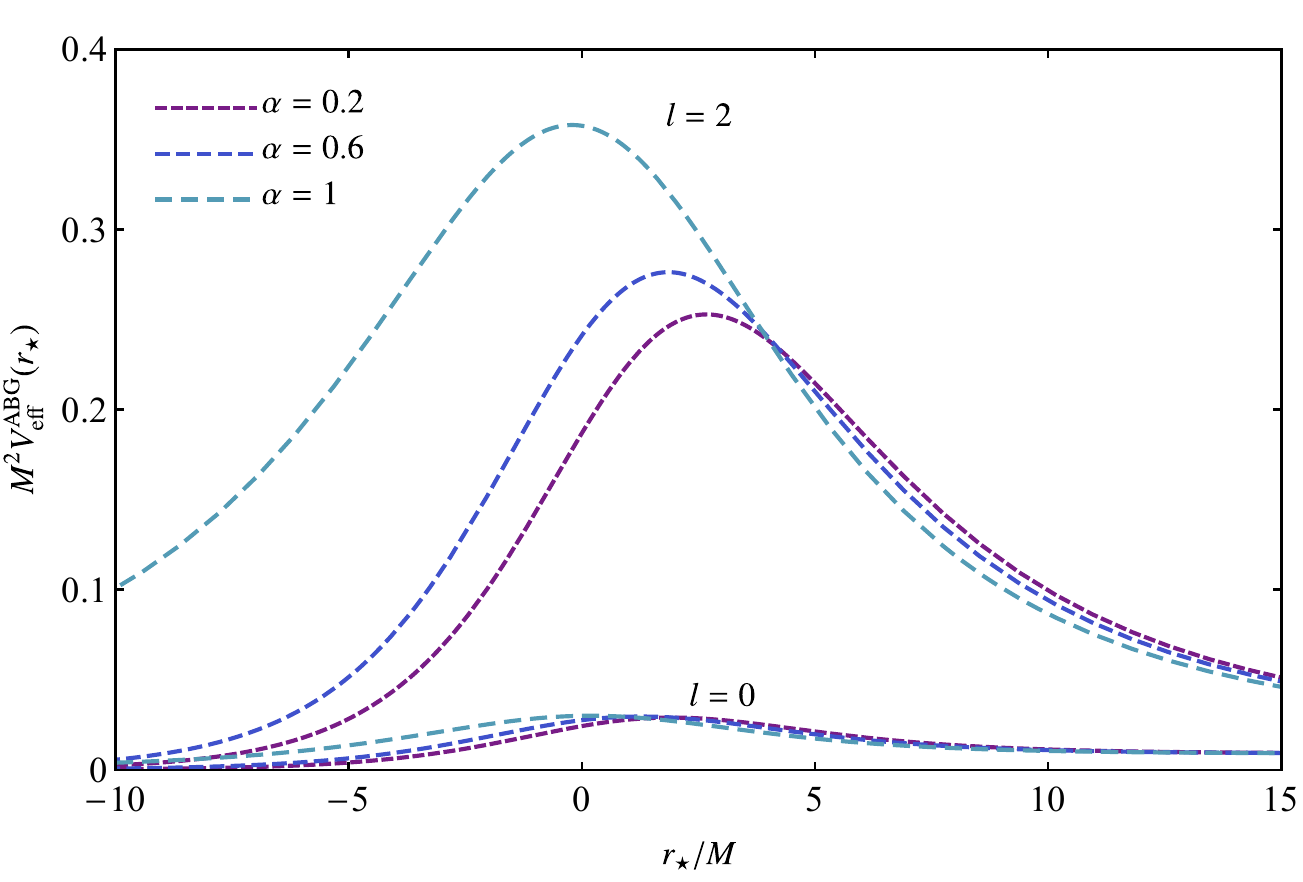}
    \includegraphics[width=1.0\columnwidth]{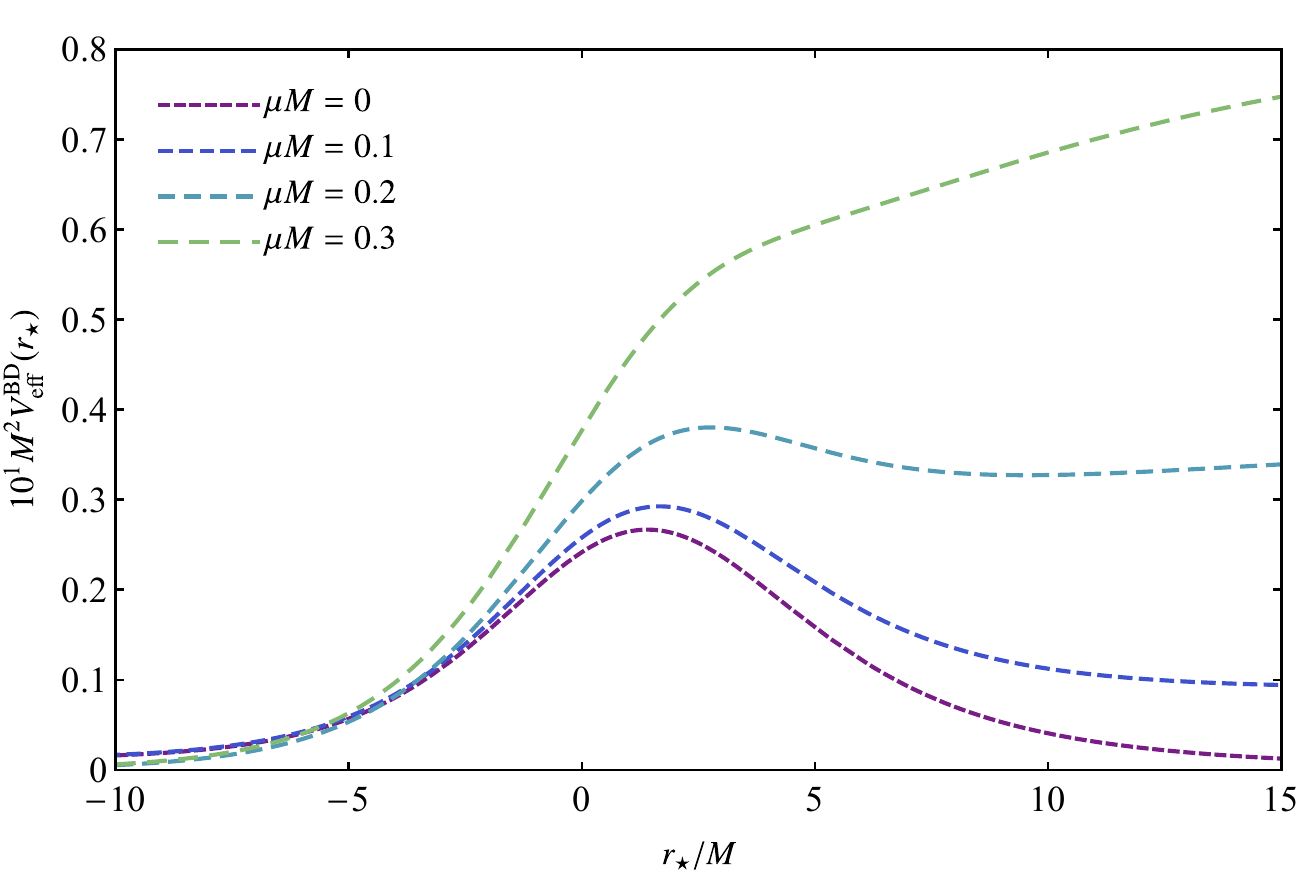}
    \caption{Effective potentials of massive scalar waves, considering two distinct scenarios: (i) ABG RBHs for distinct values of $l$ and $\alpha$, with $\mu M = 0.1$ (top panel); and (ii) BD RBHs for different choices of $\mu M$, with $l = 0$ and $\alpha = 0.4$ (bottom panel).}
    \label{peff}
\end{centering}
\end{figure}

As a function of $r_{\star}$, the asymptotic limits of $V_{\rm{eff}}(r)$ are
\begin{equation}
\label{Asym}\lim_{r_{\star} \rightarrow x} V_{\rm{eff}}(r_{\star}) = \begin{cases} 
0, & x \rightarrow - \infty,\\
\mu^{2}, & x \rightarrow +\infty .
\end{cases}
\end{equation}
In Fig.~\ref{peff}, we show the effective potentials of ABG and BD RBHs for different values of $\alpha$, $l$, and $\mu M$. We see that the peak of the effective potential increases as we increase any of these quantities. Moreover, for some values of $\mu M$, the unbounded modes are strongly absorbed [cf. $(\mu M)^{\rm{BD}} = 0.3$ case]. The effective potentials can also have a local maximum and a local minimum [cf. $(\mu M)^{\rm{BD}} = 0.2$ case]. Furthermore, we also observe that the effective potentials increase as we consider higher BH charge values. We point out that these features are similar to those observed in the RN case~\cite{CB2014}.

\begin{figure}[!htbp]
\begin{centering}
    \includegraphics[width=1.0\columnwidth]{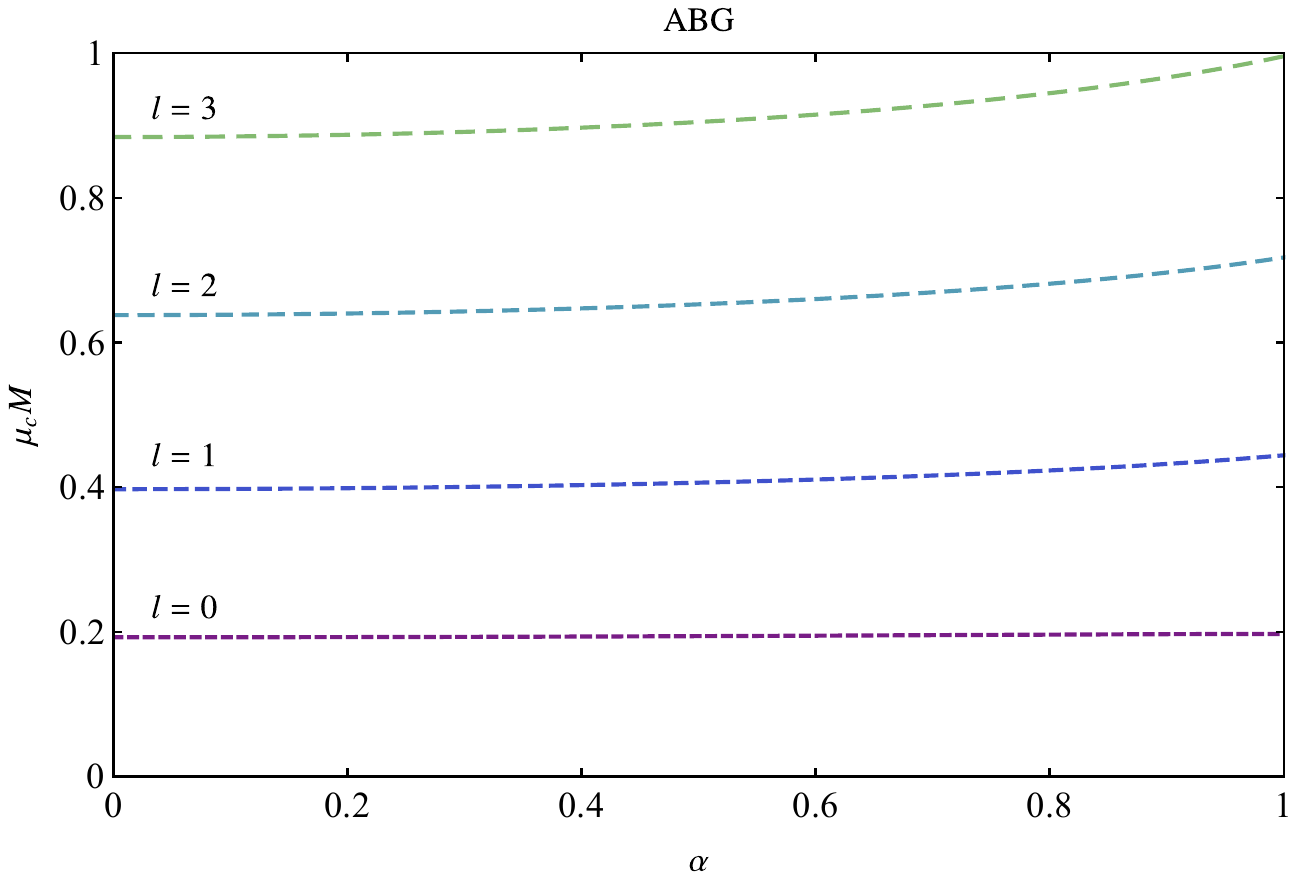}
    \caption{Critical mass of massive scalar waves in the ABG RBH spacetime, considering distinct $\alpha$ and $l$ values.}
    \label{criticalmass}
\end{centering}
\end{figure}

In fact, the values of $\mu M$ for which the unbounded modes are strongly absorbed can be mapped by introducing the notion of critical mass $\mu_{c}M$, first discussed by Jung and Park~\cite{JP2004}, which can be defined through
\begin{equation}
\label{critmass}V_{\rm{eff}}(r_{\rm{cri}}) = \mu_{c}^{2},
\end{equation}
where $r_{\rm{cri}}$ is the radial coordinate at which the effective potential achieves its local maximum. Therefore, when $\mu M > \mu_{c}M$, the scalar wave is strongly absorbed. In Fig.~\ref{criticalmass}, we show the critical mass in the ABG RBH geometry, considering distinct $\alpha$ and $l$ values. The results resemble those obtained in the RN case (see Sec. III, in particular, Fig. 2 of Ref.~\cite{CB2014}). In short, for configurations where $\mu M > \mu_{c}M$, the value of the effective potential at infinity is higher than its local maximum value, enhancing the absorption of massive scalar waves. We also notice that the results for the BD RBH are qualitatively similar to those displayed in Fig.~\ref{criticalmass}.

\begin{figure}[!htbp]
\begin{centering}
    \includegraphics[width=1.0\columnwidth]{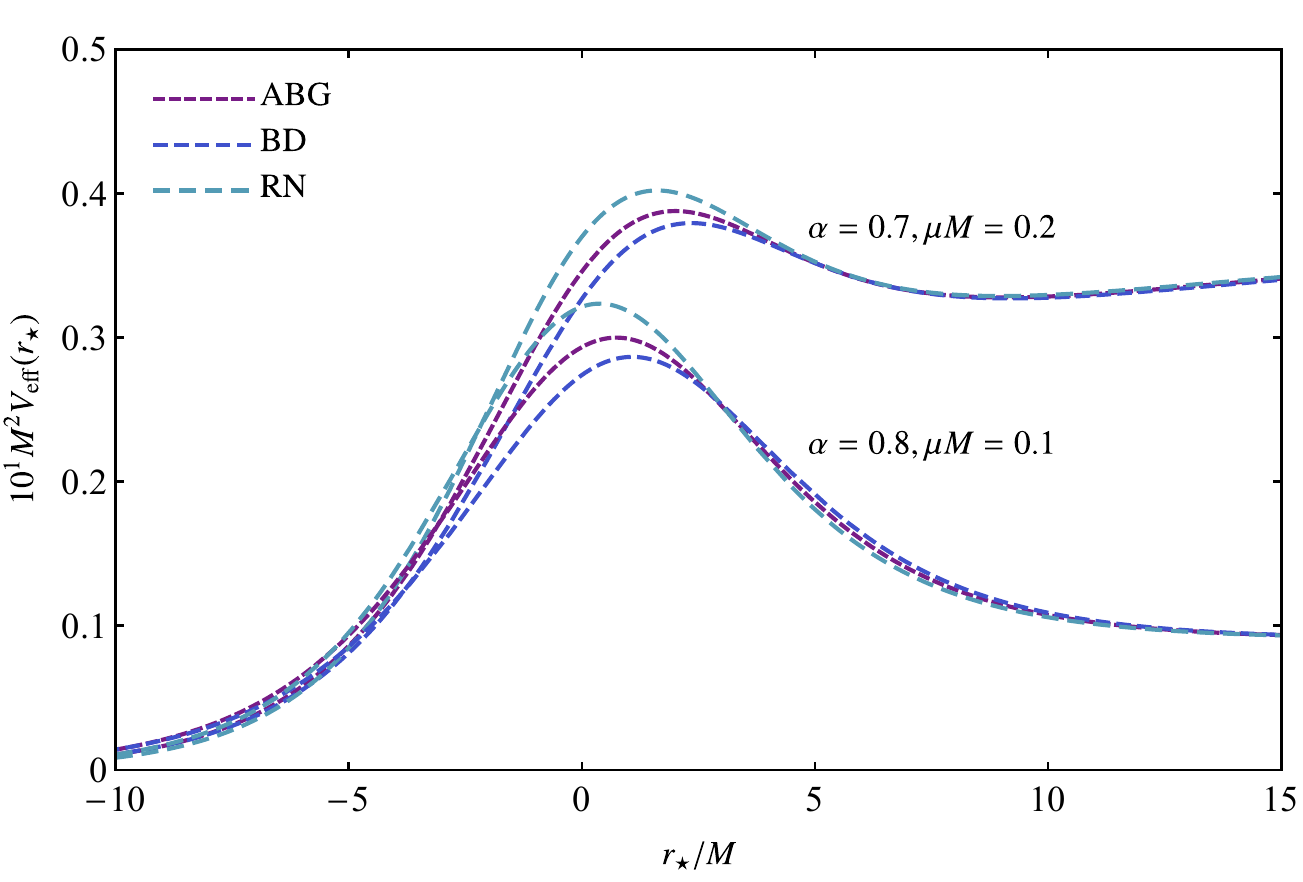}
    \caption{Comparison between the effective potentials of ABG, BD, and RN BHs. In all cases, we consider only the mode $l = 0$.}
    \label{peffabgbdrn}
\end{centering}
\end{figure}

In Fig.~\ref{peffabgbdrn}, we compare the effective potentials of ABG, BD, and RN BHs. For fixed values of $\alpha$, $l$ and $\mu M$, we see that
\begin{equation}
\label{Ratio}V_{\rm{eff}}^{\rm{RN}}\big|_{r_{\star} = r_{\rm{cri}}} > V_{\rm{eff}}^{\rm{ABG}}\big|_{r_{\star} = r_{\rm{cri}}} > V_{\rm{eff}}^{\rm{BD}}\big|_{r_{\star} = r_{\rm{cri}}}.
\end{equation}
For fixed a $\mu M$, the higher the peak of the effective potential, the lower the wave absorption. Therefore, we expect massive scalar waves to be more absorbed in the background of regular spacetimes than in that of the RN geometry. We confirm this conjecture with our numerical results for the ACS in Sec.~\ref{sec:rd}.

The boundary conditions of the Klein-Gordon equation consistent with absorption and scattering problems are given by
\begin{equation}
\label{BC}\Psi_{\omega l}\sim\begin{cases}
T_{\omega l}e^{-i\omega r_{\star}}, & r_{\star}\rightarrow -\infty  \ (r\rightarrow r_{+}),\\
e^{-i\kappa r_{\star}}+R_{\omega l}e^{i\kappa r_{\star}}, & r_{\star}\rightarrow \infty \ (r\rightarrow \infty),
\end{cases}
\end{equation}
where $T_{\omega l}$ and $R_{\omega l}$ are complex coefficients and $\kappa \equiv \sqrt{\omega^{2}-\mu^{2}}$. To have a propagating wave at infinity, the condition $\kappa > 0$, namely $\omega^{2} > \mu^{2}$, needs to be satisfied. Notice also that the condition $\kappa > 0$ implies that the parameter $v$ [cf. Eq.~\eqref{ADP}] is limited by $0 < v \leq 1$. Moreover, from the conservation of the flux, one can show that the reflection $|R_{\omega l}|^{2}$ and transmission $|T_{\omega l}|^{2}$ coefficients are related by
\begin{equation}
\label{CF}|R_{\omega l}|^{2}+\dfrac{\omega}{\kappa}|T_{\omega l}|^{2} = 1.
\end{equation}

\subsection{Absorption and differential scattering cross sections} \label{subsec:abs}

It is usual to obtain an expression for the total ACS $\sigma$ as a sum of partial wave contributions $\sigma_{l}$. For that purpose, we expand the scalar field as asymptotic plane waves and fix $C_{\omega l}$ with appropriate boundary conditions~\cite{U1976,CB2014}, yielding
\begin{equation}
\label{ACS}\sigma = \sum_{l = 0}^{\infty}\sigma_{l},
\end{equation}
where $\sigma_{l}$ is given by
\begin{equation}
\label{PACS}\sigma_{l} = \dfrac{\pi}{\kappa^{2}}(2l+1)\left(1-|R_{\omega l}|^{2}\right).
\end{equation}
In turn, the differential SCS for static and spherically symmetric spacetimes can be written as~\cite{FHM1988}
\begin{equation}
\label{SCS}\dfrac{d\sigma}{d\Omega} = |h(\theta)|^{2},
\end{equation}
where $h(\theta)$ is the scattering amplitude, given by
\begin{equation}
\label{scatta}h(\theta) = \dfrac{1}{2iv\omega}\sum_{l = 0}^{\infty}(2l+1)[e^{2i\delta_{l}(\omega)}-1]P_{l}(\cos\theta),
\end{equation}
with the phase shifts $e^{2i\delta_{l}(\omega)}$ being defined as
\begin{equation}
\label{ps}e^{2i\delta_{l}(\omega)} \equiv (-1)^{l+1}R_{\omega l}.
\end{equation}

\begin{figure}[!htbp]
\begin{centering}
    \includegraphics[width=1.0\columnwidth]{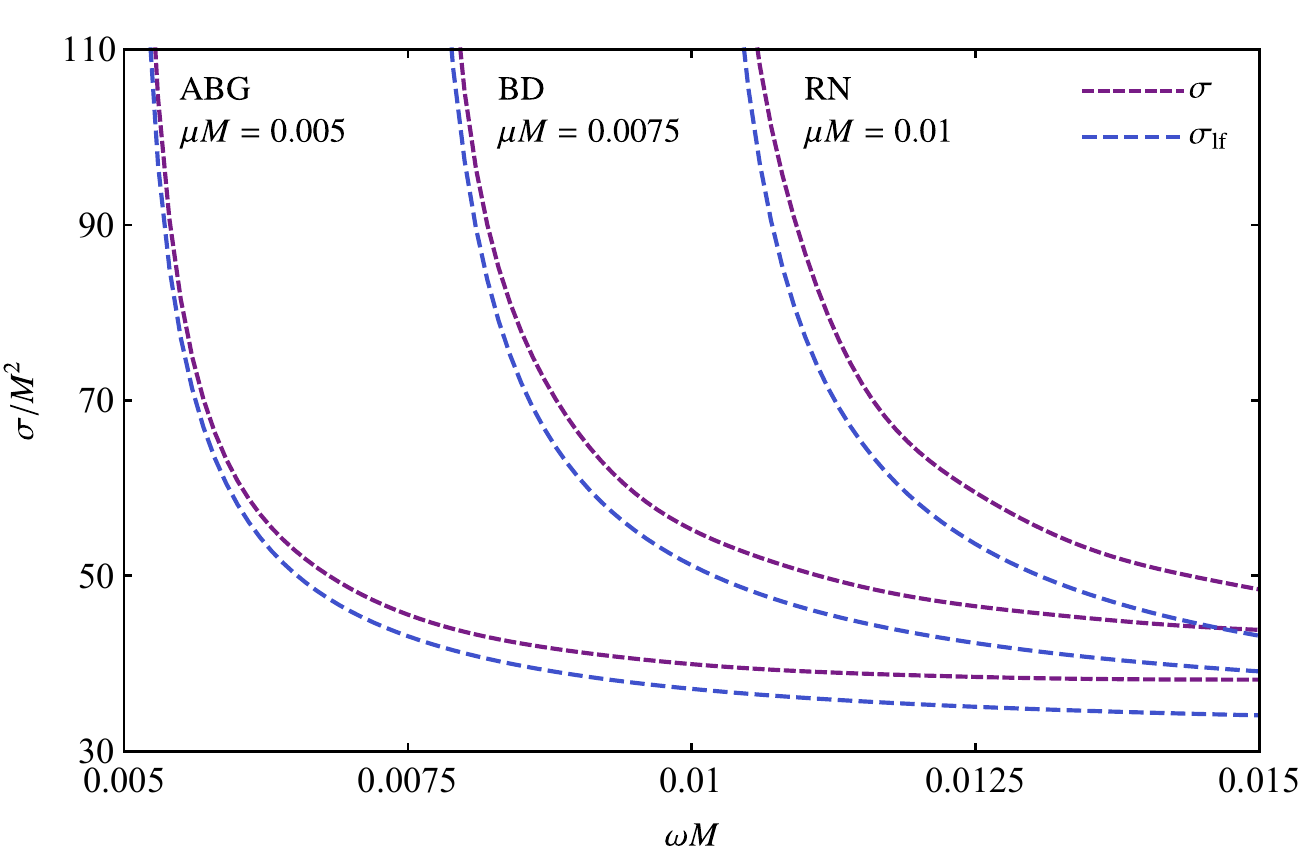}
    \caption{Comparison between the total ACSs of ABG, BD, and RN BHs with the corresponding $\sigma_{\rm{lf}}$, considering $(\mu M)^{\rm{ABG}} = 0.005$, $(\mu M)^{\rm{BD}} = 0.0075$, and $(\mu M)^{\rm{RN}} = 0.01$, respectively. We fixed $\alpha = 0.8$.}
    \label{lf2}
\end{centering}
\end{figure}

In Sec.~\ref{sec:ga}, we have discussed some analytical approximations for the absorption and scattering cross sections in the high-frequency regime. For completeness, in this section, we present an analytical approximation for the total ACS in the low-frequency regime. It was shown that, in the RN background, the very low-frequency limit of the total ACS of massive test scalar fields behaves as~\cite{CB2014}
\begin{equation}
\label{lf22}\sigma_{\rm{lf}} = \dfrac{A}{v},
\end{equation}
where $A = 4\pi r_{+}^{2}$ corresponds to the BH area. Notice that the very low-frequency limit $\omega M \ll 1$ also implies that $\mu M \ll 1$, since we are considering unbounded modes. As discussed in the addendum to Ref.~\cite{CB2014}, the approximation given by Eq.~\eqref{lf22} works well for $v \gtrsim v_{t}$, where $v_{t}$ is the transition velocity. In contrast to the critical velocity (which is obtained via analysis of the glory approximation), the transition velocity is related to the analytical approximation of the total ACS in the low-frequency regime. In Fig.~\ref{lf2}, we compare the total ACSs of ABG, BD, and RN BHs, considering the low-frequency approximation~\eqref{lf22}. Remarkably, we see that the analytical approximation obtained for the RN case works very well for the RBHs in the very low-frequency regime. We emphasize that for $v < v_{t}$, the results for the RN case cannot be generically applied to RBHs, and further clarification is needed.

\section{Results and Discussions}\label{sec:rd}

In this section, we present a selection of our main results for the absorption and scattering spectra of the ABG, BD, and RN BH spacetimes. For simplicity, we divide this section into five parts: (i) numerical method; results for (ii) absorption and (iii) scattering; (iv) a roadmap for mimicking scenarios; and (v) similar absorption and scattering properties.

\subsection{Numerical method}\label{subsec:nm}

We solve Eq.~\eqref{RE_TC} numerically from near the event horizon, i.e., $r_{\rm{ini}} = 1.0001 r_{+}$, to far away from the BH, i.e., $r_{\rm{inf}} = 10^{3}M$, aiming to find numerical values for the complex coefficients $R_{\omega l}$ and $T_{\omega l}$. This can be achieved by matching the numerical solutions of Eq.~\eqref{RE_TC}, with the appropriated boundary conditions, given by Eq.~\eqref{BC}. The linear system relating the radial function and its first derivative at the numerical infinity with the complex coefficients can be written as
\begin{align}
\label{LS}\begin{bmatrix}
\Psi_{\omega l}(r) \\ 
\Psi_{\omega l}^{\prime}(r)
\end{bmatrix} 
=
\begin{bmatrix}
e^{-i\kappa r_{\star}} & e^{i\kappa r_{\star}} \\ 
\left(e^{-i\kappa r_{\star}}\right)^{\prime} & \left(e^{i\kappa r_{\star}}\right)^{\prime}
\end{bmatrix}\cdot
\begin{bmatrix}
1/T_{\omega l} \\ 
R_{\omega l}/T_{\omega l}
 \end{bmatrix}.
\end{align}
For simplicity, we normalize the coefficients of Eq.~\eqref{BC} by $T_{\omega l}$. By solving the linear system given by Eq.~\eqref{LS}, we can find numerical values for $R_{\omega l}$ and $T_{\omega l}$, and then compute the absorption and scattering cross sections.

To obtain the differential SCS, we also adopt the convergence method developed in Refs.~\cite{YRW1954,DDL2006}, improving the results obtained from Eq.~\eqref{SCS}. This method was first introduced to improve the series convergence in the context of the Coulomb scattering~\cite{YRW1954}. Basically, given a Legendre polynomial series, e.g.,
\begin{equation}
g(\theta) = \sum_{l = 0}a_{l}^{0}P_{l}(\cos\theta),
\end{equation}
which is divergent at $\theta = 0$, it is possible to define a mth reduced series, given by
\begin{equation}
(1-\cos\theta)^{m}g(\theta) = \sum_{l = 0}a_{l}^{(m)}P_{l}(\cos\theta).
\end{equation}
The reduced series is less divergent at $\theta = 0$, so it is expected to converge faster. Using the properties of the Legendre polynomials~\cite{AS1965}, we can show that the new coefficients $a_{l}^{(i+1)}$ are related to the old coefficients $a_{l}^{i}$, with $i$ being the iteration index, by the following iterative formula:
\begin{equation}
a_{l}^{(i+1)} = a_{l}^{(i)}-\dfrac{l+1}{2l+3}a_{l+1}^{(i)} - \dfrac{l}{2l-1}a_{l-1}^{(i)}.
\end{equation}
Dolan \textit{et al.} adapted this method for the scattering of compact objects~\cite{DDL2006}. For our purposes, two or three iterations are sufficient to sum the series numerically with an excellent precision. Further details on the series convergence can be found in Ref.~\cite{DS2017}. Notice also that the absorption and scattering spectra strongly depends on the summations in the angular momentum $l$. We typically set $l_{\rm{max}} = 10$, for absorption problems, and $l_{\rm{max}} = 50$, for scattering problems.

\begin{figure}[!htbp]
\begin{centering}
    \includegraphics[width=1.0\columnwidth]{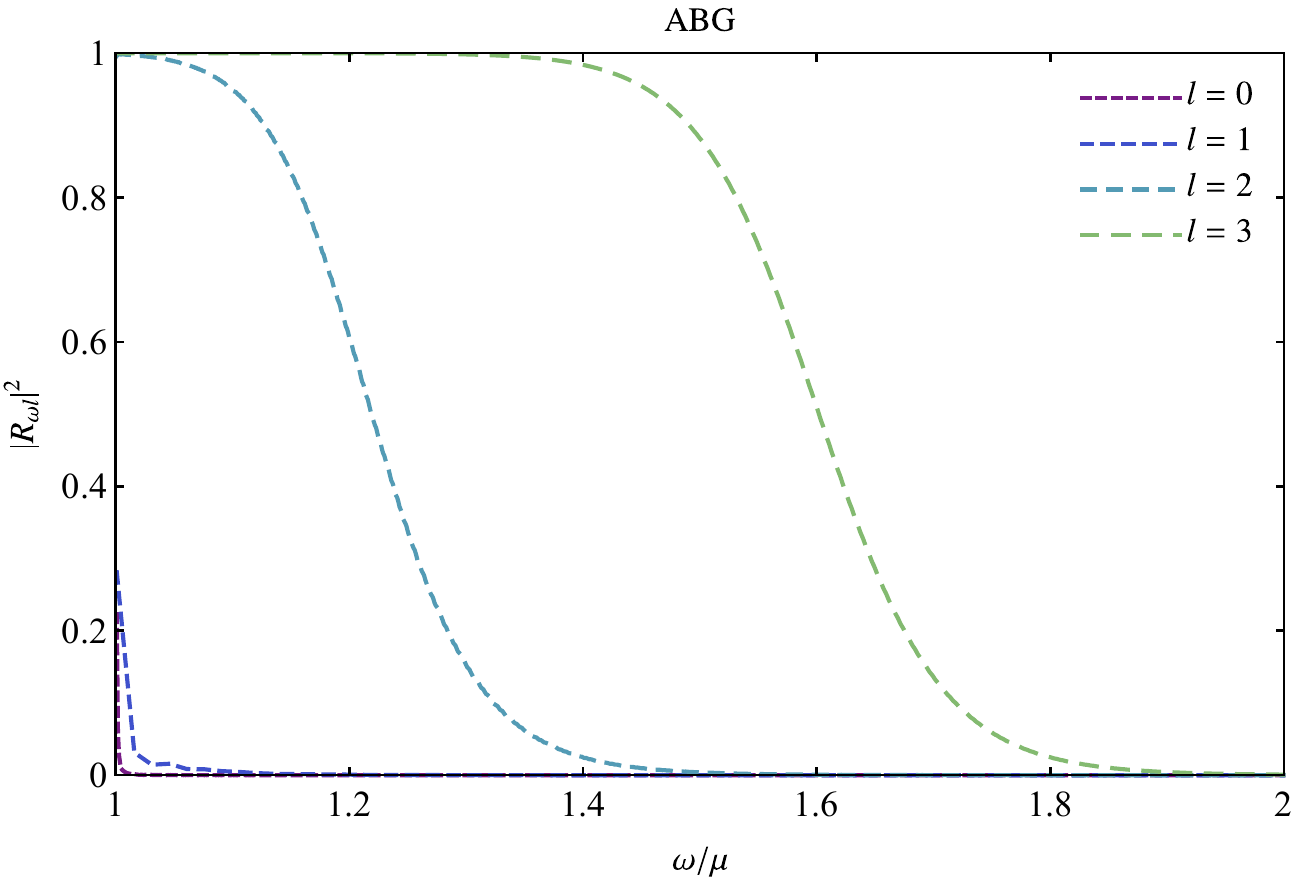}
    \includegraphics[width=1.0\columnwidth]{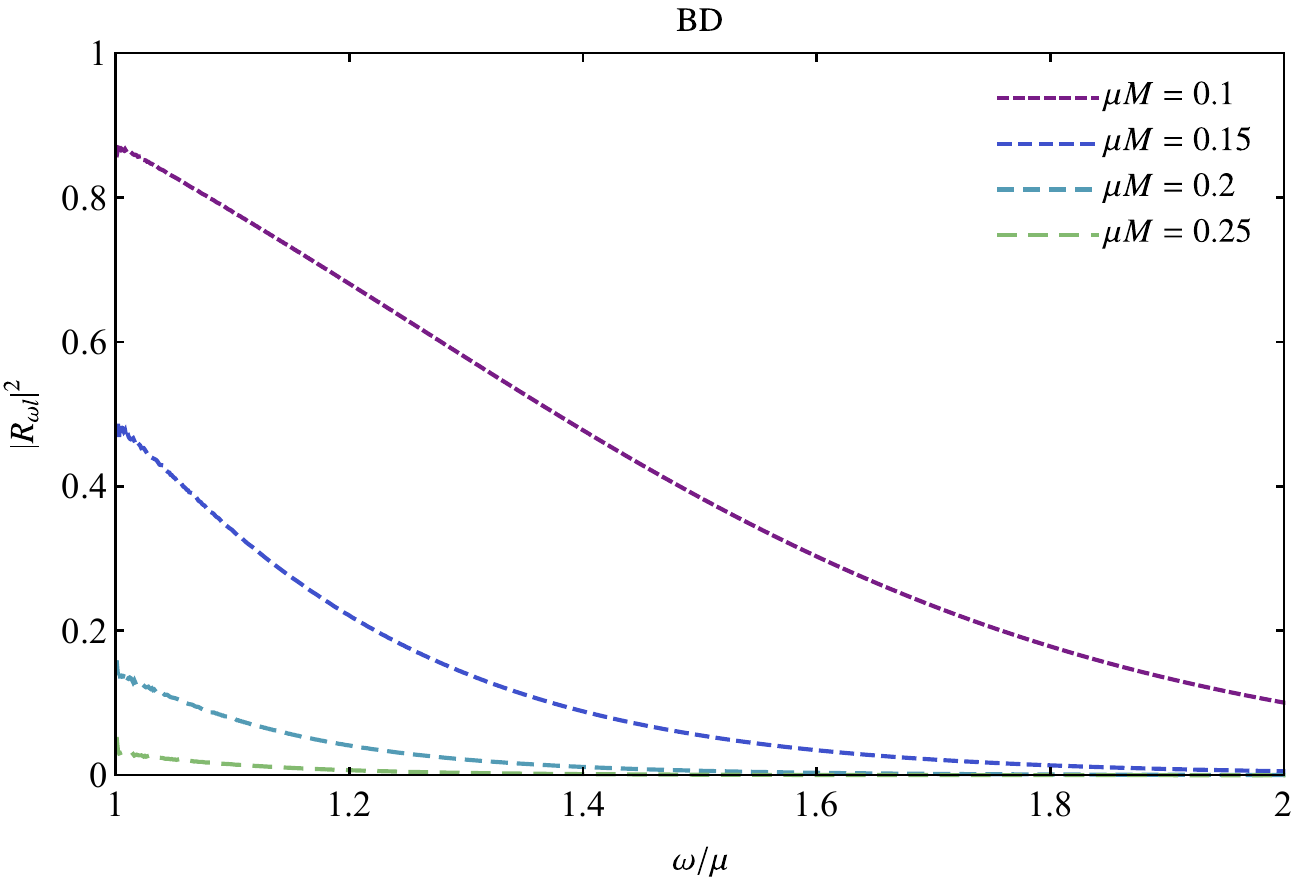}
    \caption{Reflection coefficients of the massive scalar wave, considering distinct scenarios, namely: (i) different values of $l$ for $\mu M = 0.5$ in the ABG background (top panel); and (ii) different choices of $\mu M$ for $l = 0$ in the BD geometry (bottom panel). In both scenarios, we consider $\alpha = 0.8$.}
    \label{TRA&REFABGBDlQ}
\end{centering}
\end{figure}

In Fig.~\ref{TRA&REFABGBDlQ}, we show examples of reflection coefficients for ABG and BD RBHs. In the top panel, we notice that the reflection coefficient for the first mode ($l = 0$) is almost null. This may be understood by noting that, for this situation, the critical mass is given by $(\mu_{c}M)^{\rm{ABG}} \approx 0.1961$ and $\mu M > (\mu_{c}M)^{\rm{ABG}}$. In this context, we expect the $l = 0$ mode to be strongly absorbed. In the bottom panel, we observe a similar effect, but now considering distinct values of $\mu M$ in the BD geometry. For this case, as we consider higher $\mu M$ values, we observe nearly total absorption when $\mu M >(\mu_{c}M)^{\rm{BD}} \approx 0.1917$. Furthermore, we observe that, for modes $l = 0$ and $l = 1$, the reflection coefficient does not approach unity in the low-frequency regime. This contrasts with the massless scalar field case.

\begin{figure}[!htbp]
\begin{centering}
    \includegraphics[width=1.0\columnwidth]{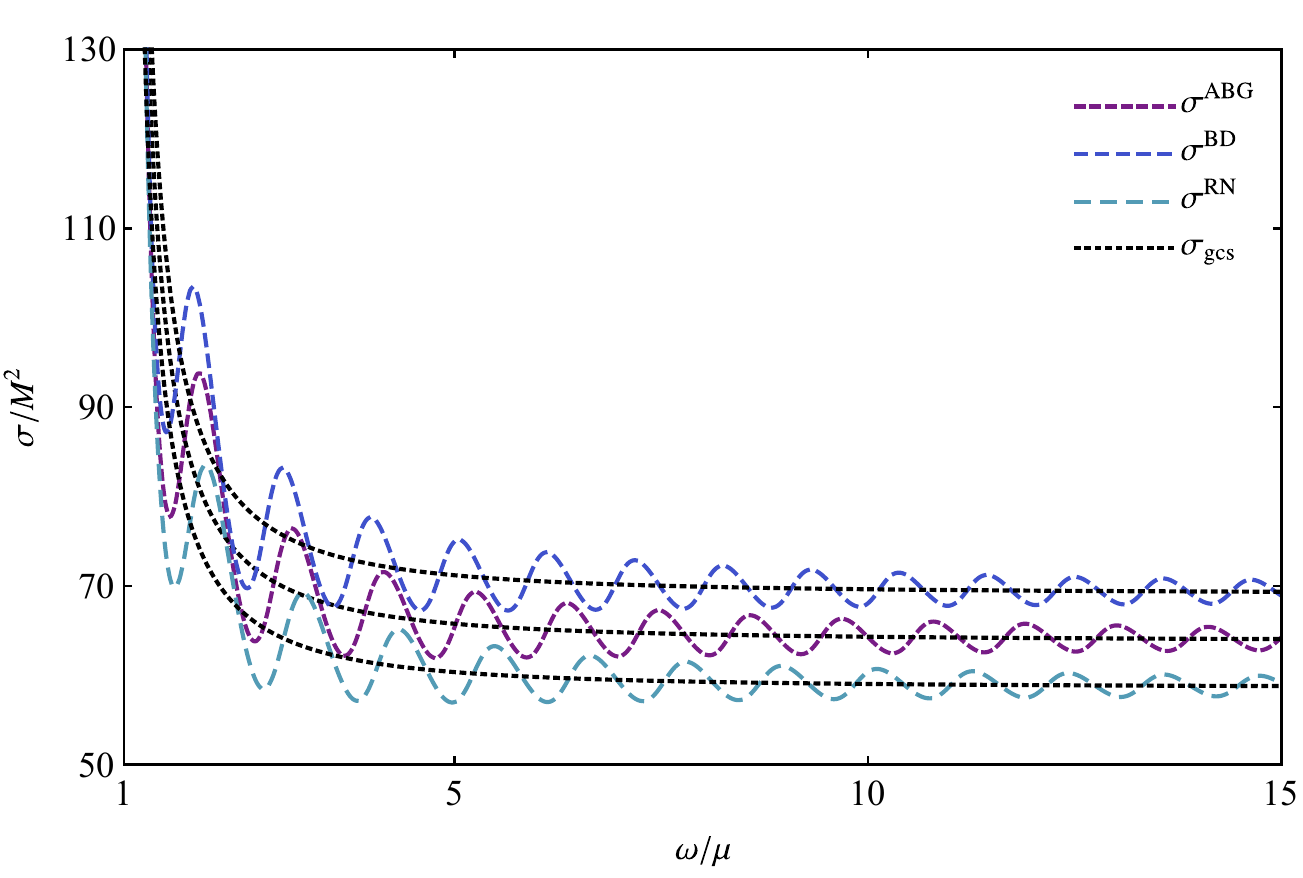}
    \includegraphics[width=1.0\columnwidth]{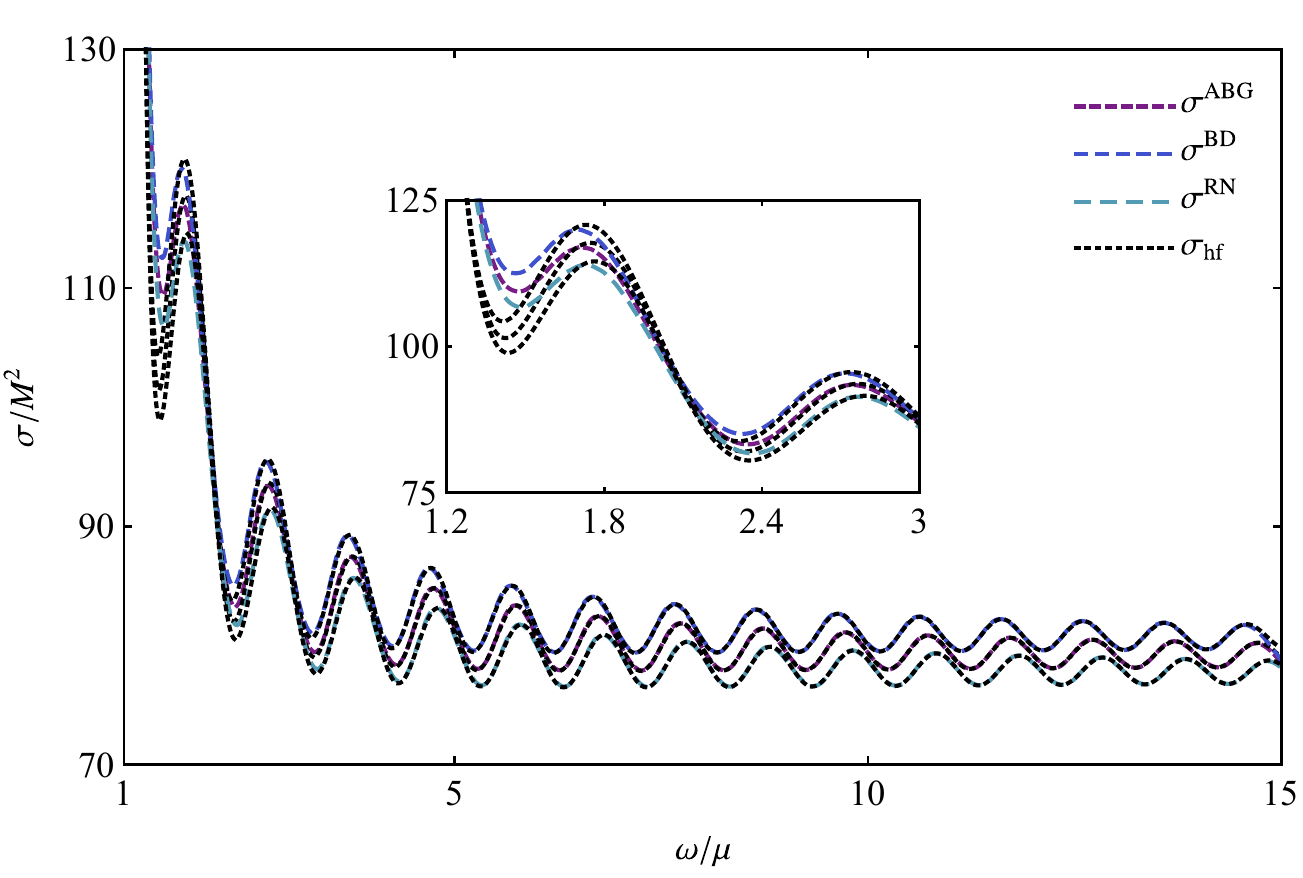}
    \caption{Total ACSs of ABG, BD, and RN BHs, considering the corresponding: (i) GCSs for $\alpha = 0.9$ (top panel); and (ii) sinc approximations for $\alpha = 0.5$ (bottom panel). We consider $\mu M = 0.2$. The inset in the bottom panel shows the cross section in a given frequency range to better visualize the similarity.}
    \label{GCSSINC}
\end{centering}
\end{figure}

In Fig.~\ref{GCSSINC}, we exhibit a selection of our numerical results computed in the whole frequency regime with the corresponding classical $\sigma_{\rm{gcs}}$ and semiclassical $\sigma_{\rm{hf}}$ approximations in the high-frequency limit. We note that the total ACS oscillates around the corresponding GCS in the high-frequency regime, and the oscillatory pattern is well described by the sinc approximation, even for moderate frequency values. Moreover, the total ACSs of the BHs decrease as we increase $\alpha$. This is expected from the behavior of the effective potential peak, which increases as we consider higher values of BH charge.

\begin{figure}[!htbp]
\begin{centering}
    \includegraphics[width=1.0\columnwidth]{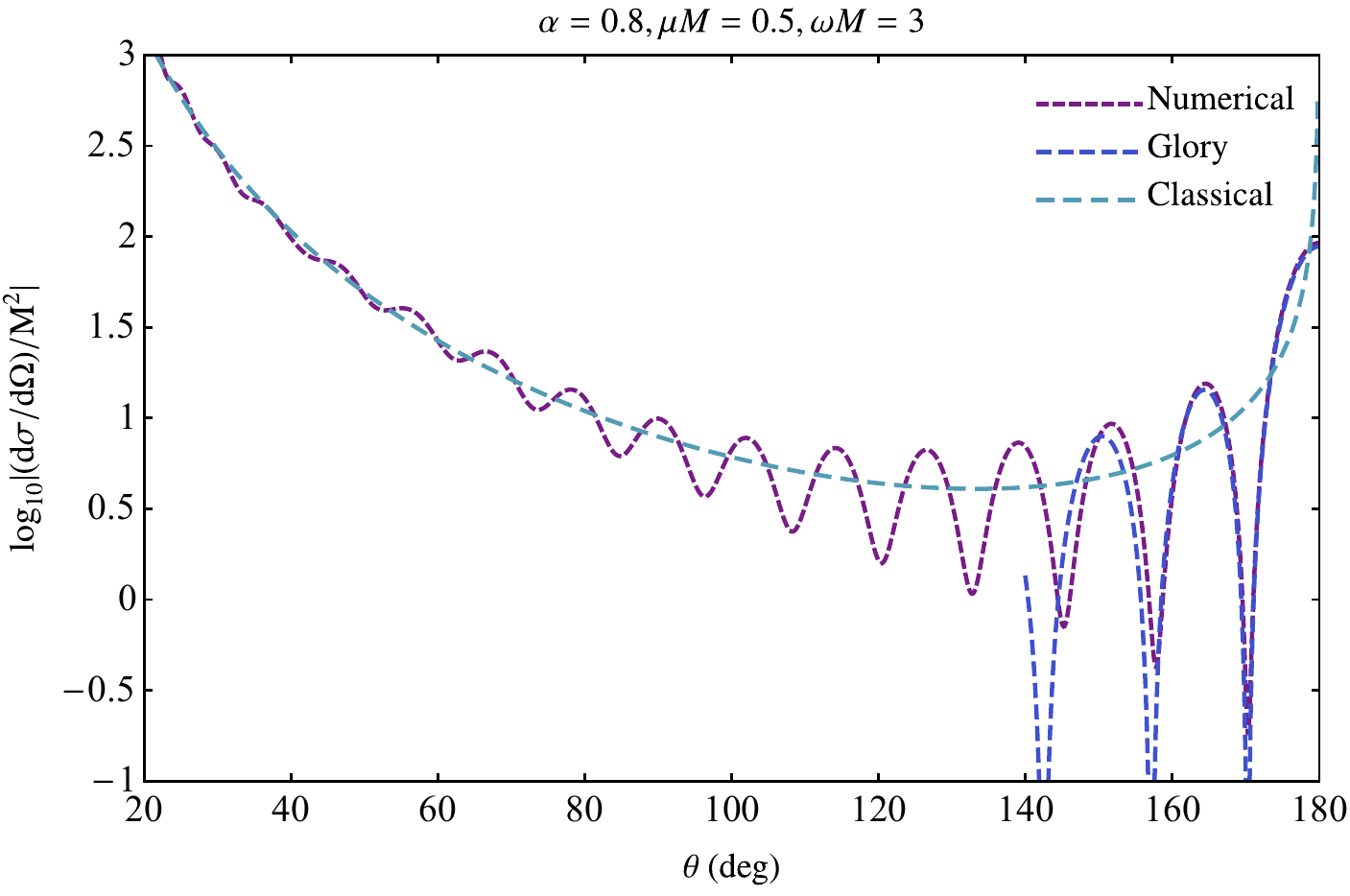}
    \caption{Comparison of the differential SCS of the RN BH obtained numerically with the corresponding differential classical SCS and glory approximation, for $\omega M = 3$, $\mu M = 0.5$, and $\alpha = 0.8$.}
    \label{CSCSABGBDRN}
\end{centering}
\end{figure}

In Fig.~\ref{CSCSABGBDRN}, we compare the differential SCS of massive scalar fields in the background of the RN BH obtained numerically with the corresponding classical differential SCS derived analytically, and we also exhibit the corresponding glory approximation.  We see that the  differential SCS diverges as we approach the forward direction $(\theta \rightarrow 0^{\circ})$, as anticipated from the behavior of the differential classical SCS in the weak field limit [cf. Eq.~\eqref{CSCSweakABG}]. Moreover, the semiclassical glory [cf. Eq.~\eqref{glory}] captures very well the oscillatory pattern near the backward direction $(\theta \rightarrow 180^{\circ})$. The numerical results for ABG and BD RBHs exhibit a similar pattern, and we chose not to show them explicitly here. The results presented in Fig.~\ref{lf2}, combined with those obtained in this section, show that the analytical and numerical results for the absorption and scattering cross sections agree very well in their corresponding limits. We also point out that the oscillatory pattern in the absorption and scattering spectra is associated with the interference of the scalar waves propagating around the BH.

\subsection{Absorption results}\label{subsec:absABGandBD}

\begin{figure}[!htbp]
\begin{centering}
	\includegraphics[width=1.0\columnwidth]{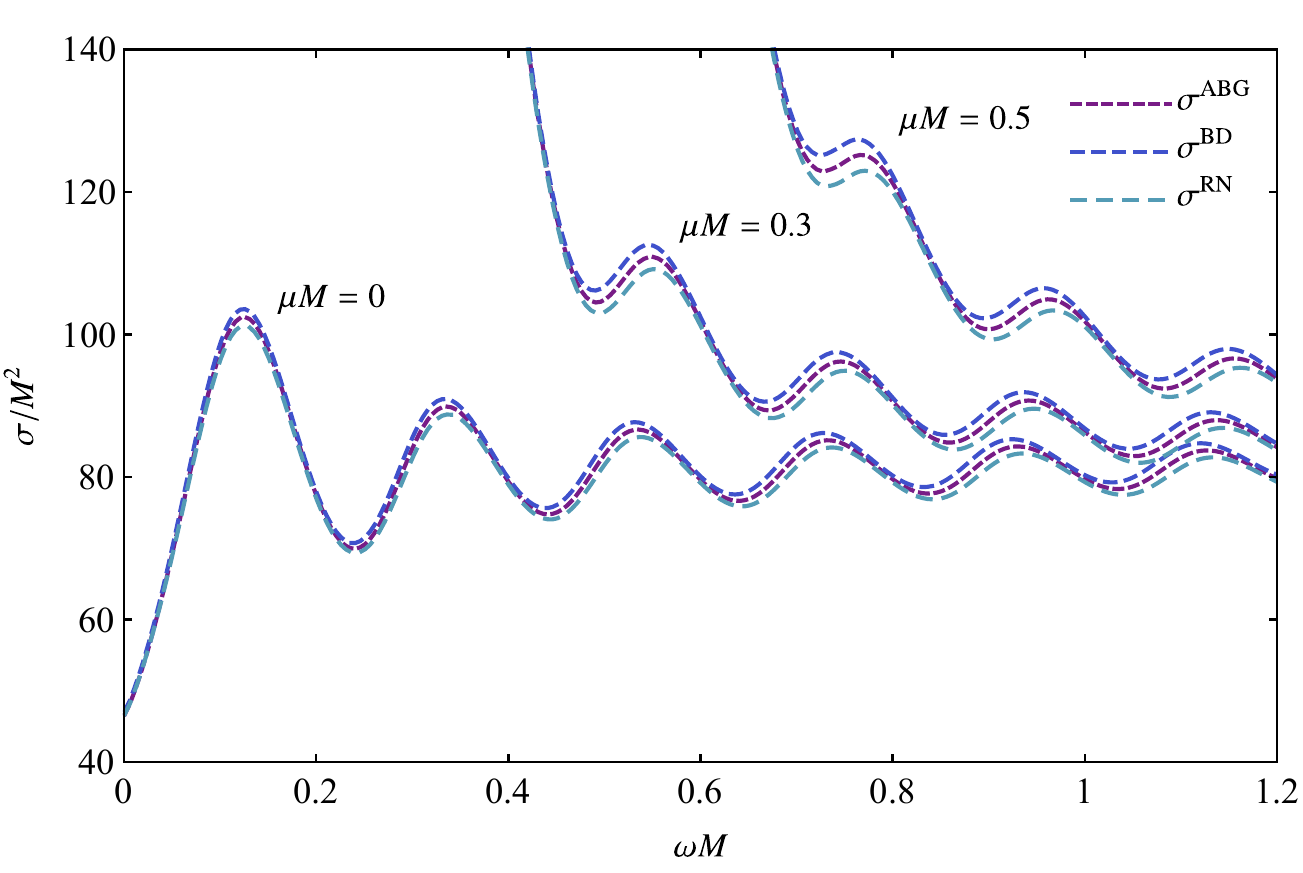}
    \caption{Total ACSs of massive scalar waves in the background of ABG, BD, and RN BHs, considering distinct values of $\mu M$, with $\alpha = 0.4$. For the massless case, the ACSs of the BHs in the limit $\omega M \rightarrow \mu M$ tend to $\sigma^{\rm{ABG}} \approx 46.13M^2$, $\sigma^{\rm{BD}} \approx 46.59M^2$, and $\sigma^{\rm{RN}} \approx 46.15M^2$, which are the corresponding BH areas.}
    \label{TACSABGBDRNQm}
\end{centering}
\end{figure}

In Fig.~\ref{TACSABGBDRNQm}, we display the total ACSs of massive scalar waves in ABG, BD, and RN BH spacetimes for different values of $\mu M$. Generically, besides the similarity at $\omega M \rightarrow \mu M$, the total ACSs of the BHs enhance as we increase $\mu M$ and diverge in the limit $\omega M \rightarrow \mu M$. Moreover, we observe that, for fixed values of $\alpha$ and $\mu M$, the total ACSs of ABG, BD, and RN BHs typically satisfy
\begin{equation}
\label{tacsabgbdrn}\sigma^{\rm{BD}} > \sigma^{\rm{ABG}} > \sigma^{\rm{RN}}.
\end{equation}
Therefore, the scalar waves are more absorbed in the background of ABG and BD RBHs than in the RN BH, which is compatible with the behavior of the potential barrier (cf. Sec.~\ref{sec:sf}, in particular, Fig.~\ref{peffabgbdrn}). In the low-frequency regime, we also notice that for $\mu M = 0$, the ACS tends to the surface area of the BH, as expected (see, e.g., Refs.~\cite{CDE2009,MC2014,PLC2020,DGM1997,H2001}).

\begin{figure}[!htbp]
\begin{centering}
    \includegraphics[width=1.0\columnwidth]{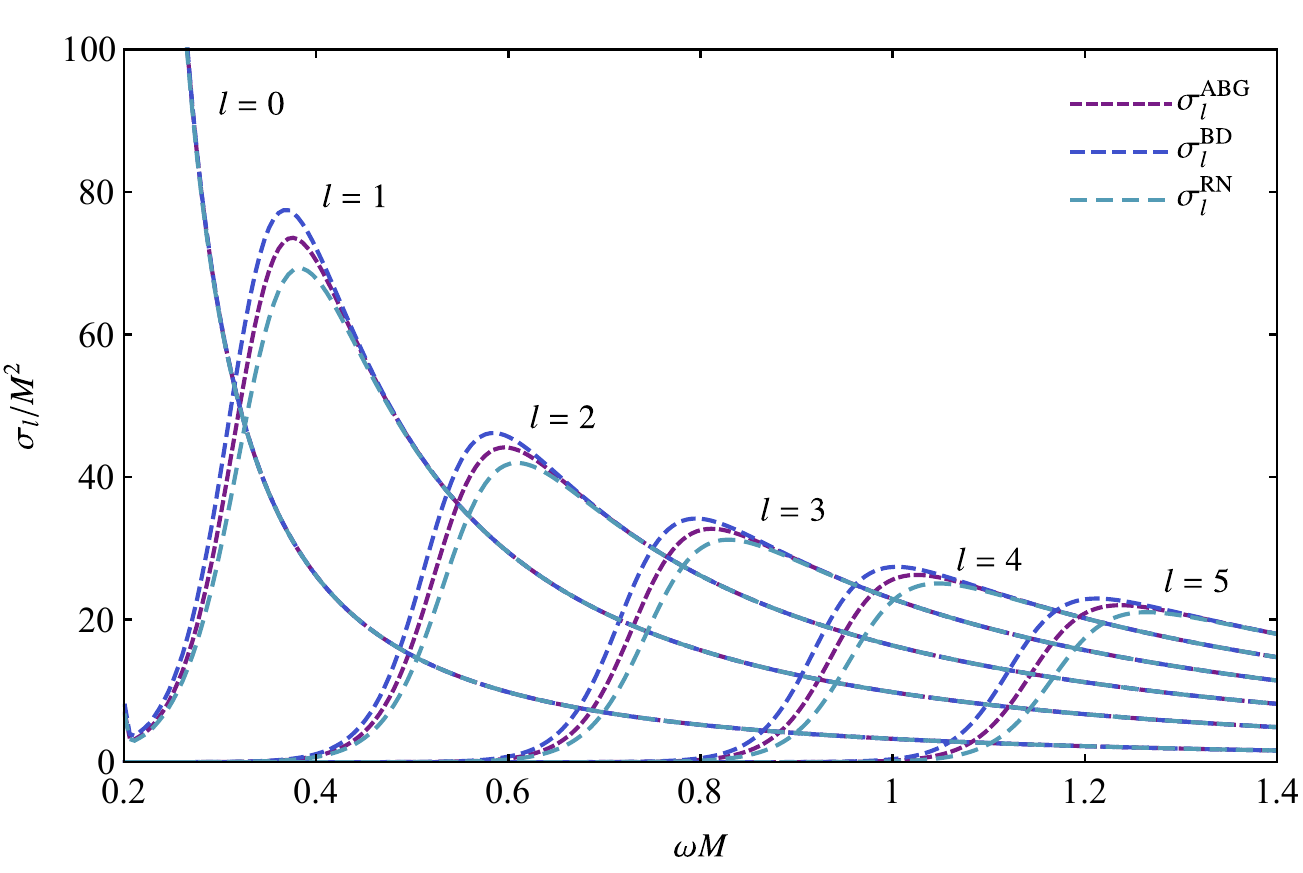}
    \caption{Partial ACSs of ABG, BD, and RN BHs, considering different values of $l$ and fixed values of $\alpha = 0.7$ and $\mu M = 0.2$.}
    \label{PACSABGBDRN}
\end{centering}
\end{figure}

In Fig.~\ref{PACSABGBDRN}, we analyze the partial ACSs of ABG, BD, and RN BHs for different values of $l$. The peak of the partial ACS decreases as we consider higher values of $\omega M$. We notice that as $\omega M \rightarrow \mu M$, the mode $l = 1$ also contributes to the ACS (see also Fig.~\ref{TRA&REFABGBDlQ}), which differs from the massless case, where the regime $\omega M \rightarrow 0$ is entirely described by $\sigma_{0}$. Hence, for the massive case, the fundamental mode is not necessarily the whole contribution for the total ACS in the low-frequency regime. Moreover, we also observe that the peak of the partial ACSs typically satisfies
\begin{equation}
\sigma^{\rm{BD}}_{l} > \sigma^{\rm{ABG}}_{l} > \sigma^{\rm{RN}}_{l},
\end{equation}
in agreement with Eq.~\eqref{tacsabgbdrn}.

\subsection{Scattering results}\label{subsec:scattABGandBD}

\begin{figure}[!htbp]
\begin{centering}
    \includegraphics[width=1.0\columnwidth]{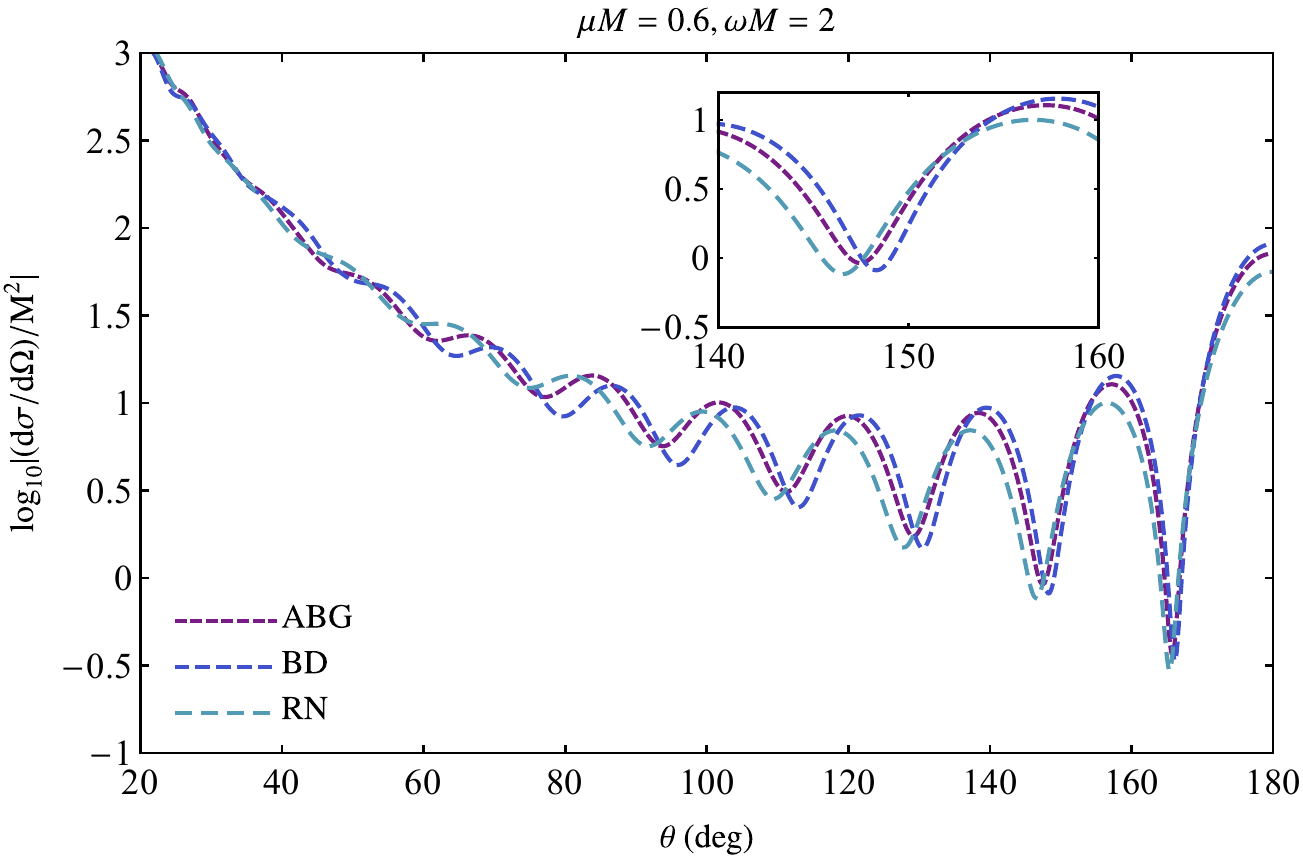}
    \includegraphics[width=1.0\columnwidth]{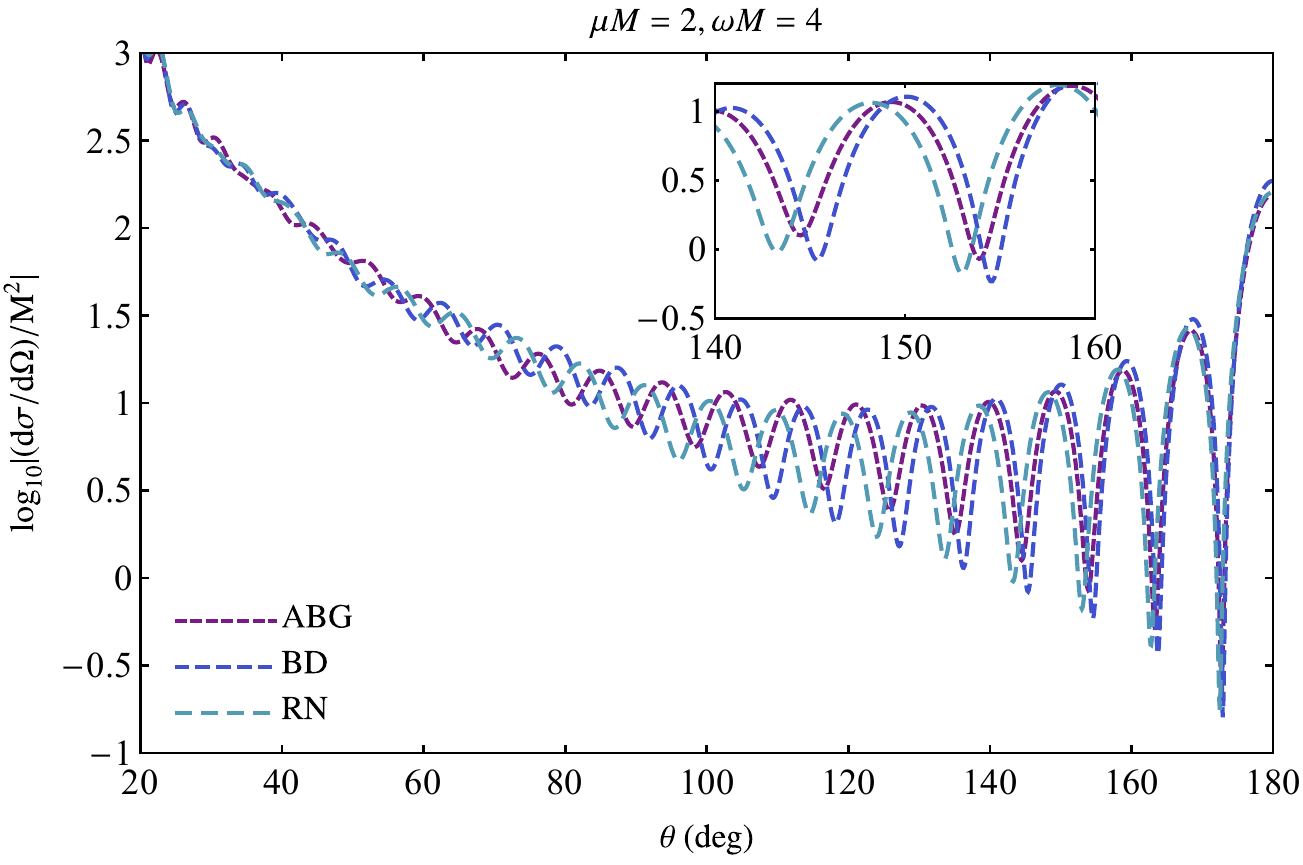}
    \caption{Comparison of the differential SCSs of ABG, BD, and RN BHs, considering: (i) $\mu M = 0.6$ and $\omega M = 2$ (top panel); and (ii) $\mu M = 2$ and $\omega M = 4$ (bottom panel). We fixed  $\alpha = 0.8$. The insets zoomed the differential SCSs for a given range of the scattering angle.}
    \label{TSCSABGBDRNome}
\end{centering}
\end{figure}
In Fig.~\ref{TSCSABGBDRNome}, we compare the differential SCSs of ABG, BD, and RN BHs. We note that the interference fringe widths of the RN BH are larger than that of the ABG RBH, which, in turn, are larger than that of the BD RBH. We also notice that the interference fringe widths decrease as we consider higher values of the frequency, for fixed values of BH charge and field mass. This behavior can also be related to the glory formula [cf. Eq.~\eqref{glory}] because the interference fringe width is inversely proportional to the frequency, as can be noticed from the argument of the Bessel function.

\begin{figure*}[!htbp]
\begin{centering}
    \includegraphics[width=1.0\columnwidth]{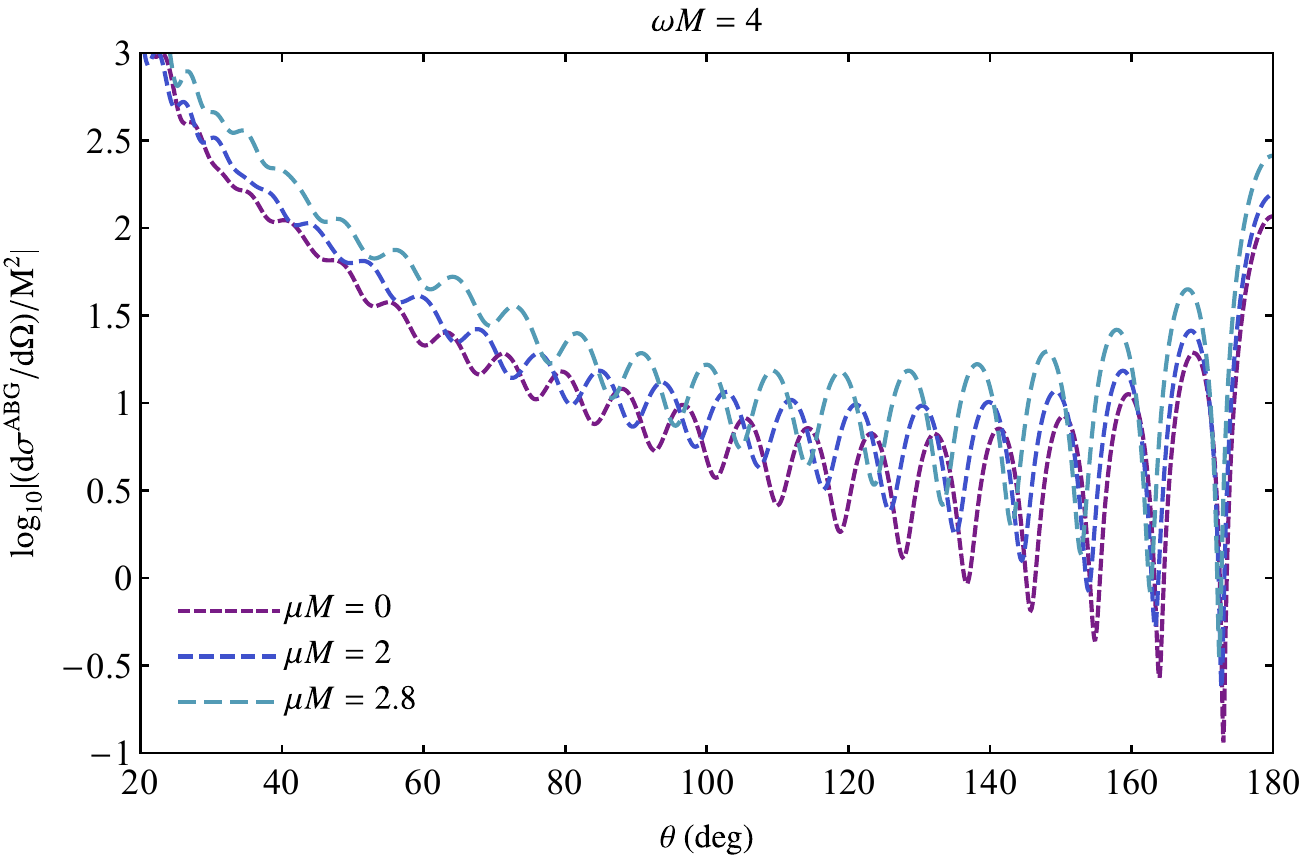}
   	\includegraphics[width=1.0\columnwidth]{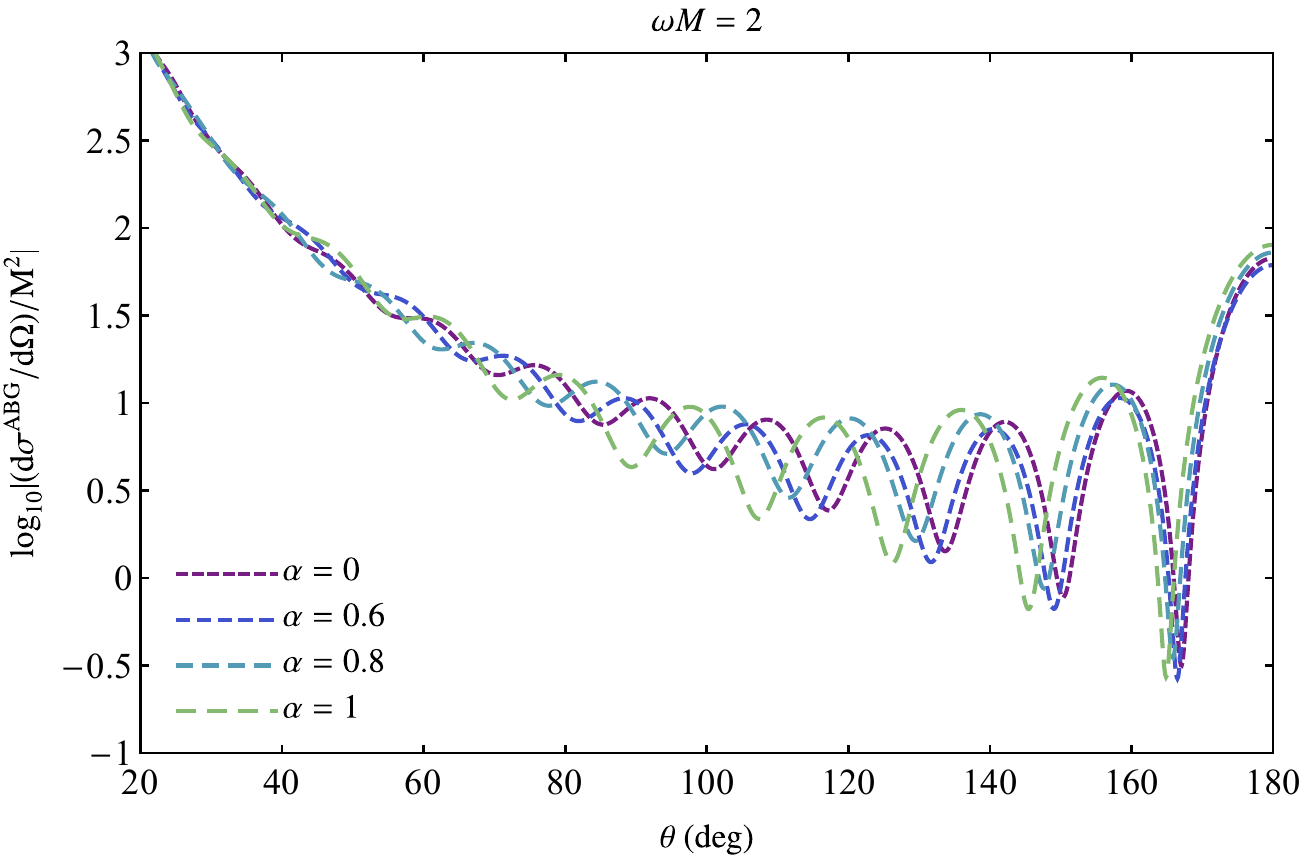}
    \includegraphics[width=1.0\columnwidth]{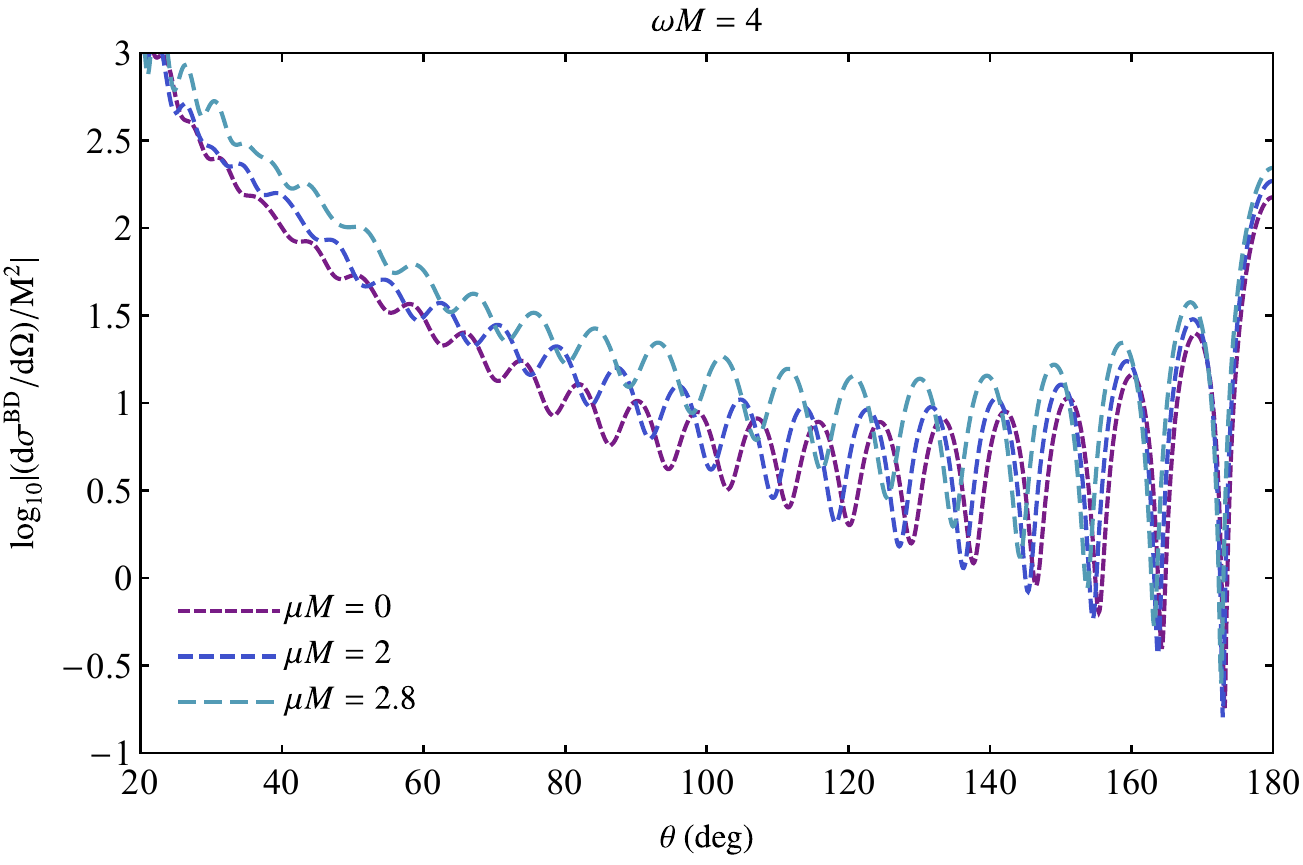}
    \includegraphics[width=1.0\columnwidth]{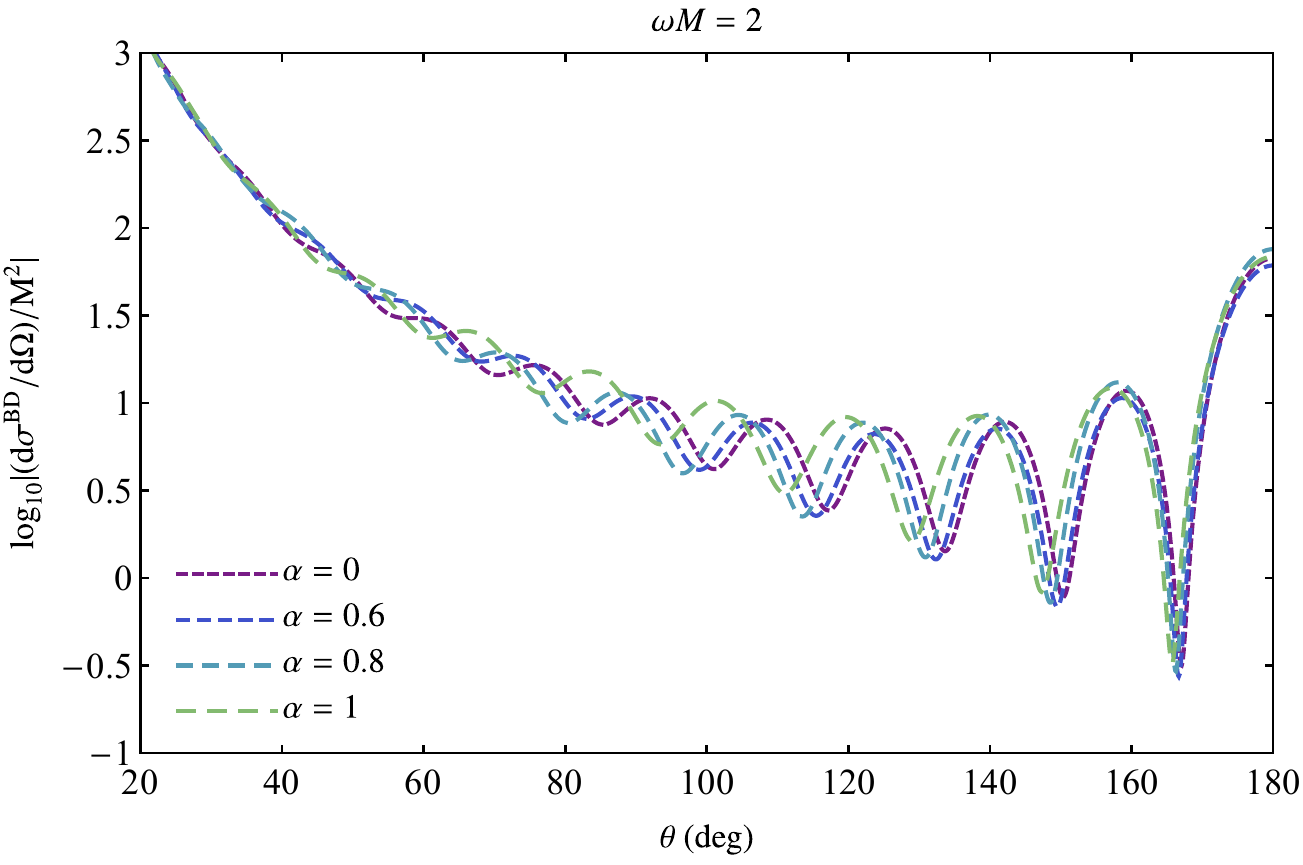}
    \includegraphics[width=1.0\columnwidth]{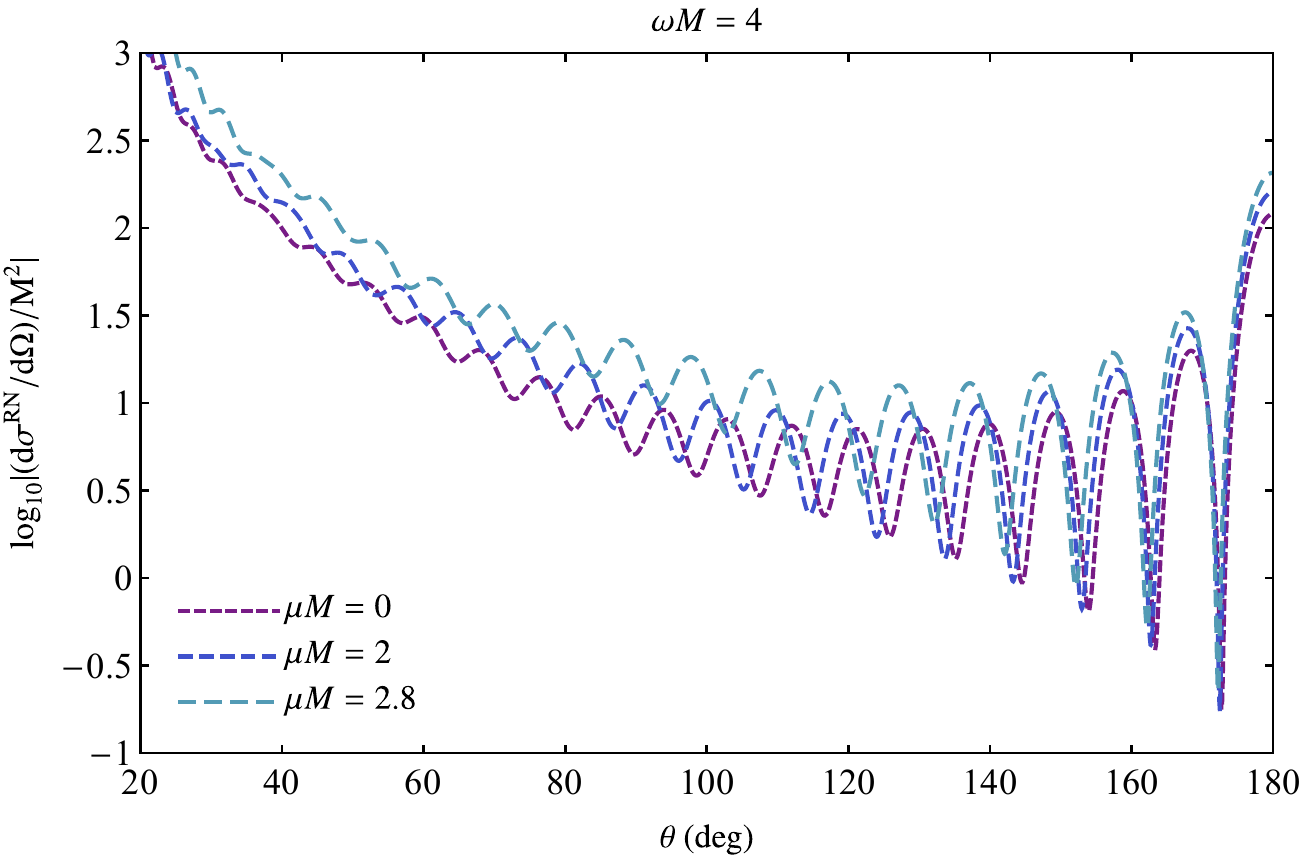}
    \includegraphics[width=1.0\columnwidth]{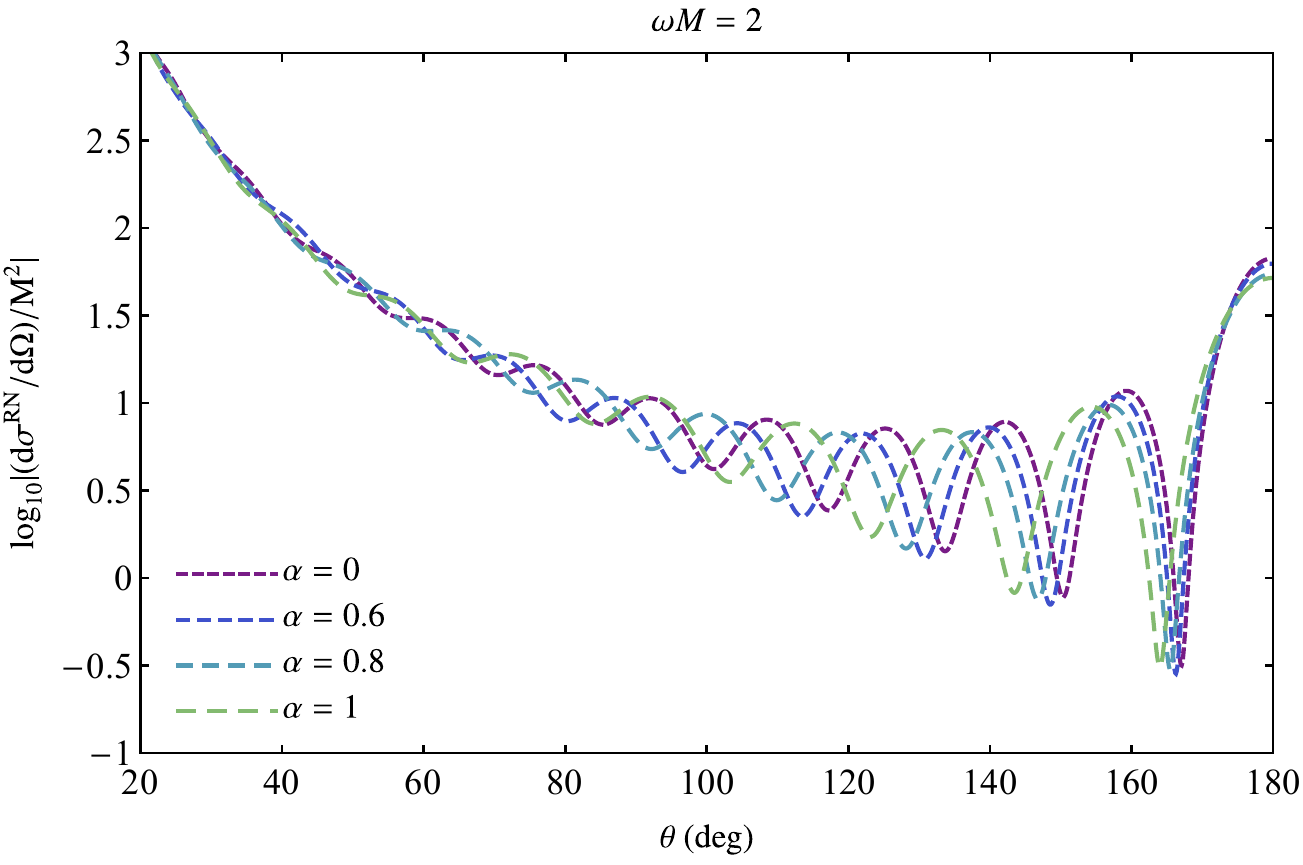}
    \caption{Differential SCSs of ABG (top panels), BD (middle panels), and RN (bottom panels) BHs, considering: (i) different values of $\mu M$, with $\omega M = 4$ and $\alpha = 0.8$ (left panels); and (ii) distinct choices of $\alpha$, with $\omega M = 2$ and $\mu M = 0.4$ (right panels).}
    \label{TSCSABGBDRN}
\end{centering}
\end{figure*}

The differential SCSs of ABG, BD, and RN BHs for different values of the mass of the field and charge of the BH are investigated in Fig.~\ref{TSCSABGBDRN}. We see that the interference fringe widths increase (i) as we consider higher values of the field mass coupling, for fixed $\alpha$, and (ii) as we consider higher BH charge-to-mass ratios, for fixed $\omega M$. This behavior can be related to the behavior of the glory formula~\eqref{glory}. According to it, near the backward direction, the widths are inversely proportional to $v b_{g}$, and, therefore, increase as $vb_{g}$ decreases for $v > v_{c}$. We notice that, for all the combinations used in Fig.~\ref{TSCSABGBDRN} in the absorption and scattering parameter space, we have $v > v_{c}$. We will present a detailed analysis considering $v \leq v_{c}$ elsewhere. Moreover, the flux intensity typically increases as we consider higher values of $\mu M$. However, as we vary the BH charge-to-mass ratio, the intensity flux profile differs for each BH.

\subsection{Similar cross sections: A roadmap}

Since RBHs stand out as potential alternatives to singular BHs in standard GR, it is interesting to verify whether observable quantities, such as the ACS, of singular BHs can be mimicked by RBHs. In particular, for NED-sourced RBH solutions, this is relevant as this class of RBHs can have a BH geometry very similar to standard BHs, but with distinct features from the phenomenological point of view. For example, photons in NED follow null geodesics of an effective geometry~\cite{GDP1981,MN2000} and, in certain geometries, we may have unbounded superradiance~\cite{MP2024b}, superradiant instability~\cite{MAP2024c}, and scalar clouds~\cite{S2024} in spherically symmetric NED-sourced spacetimes. Therefore, search for configurations in which singular and regular BHs lead to the same observable quantities allows to better understand the characteristic signatures of alternative BH solutions, and in which scenarios these characteristics are most likely to be relevant.

We will first discuss how the field mass contributes to the similar cross sections based on the analytical approximations. We begin, however, revisiting the results for the massless scalar field case. In this case, the similarity of the total ACS in the whole frequency range between regular and singular BHs can be understood by investigating the corresponding BH area of the pairs considered. The percentage deviation between the BH area of each pair $(\alpha^{\rm{RBH}},\alpha^{\rm{RN}})$ can be defined as 
\begin{equation}
\label{EHD_PCT}\mathrm{A}_{\rm{dif}} \equiv \dfrac{\left|\mathrm{A}^{\rm{RN}} - \mathrm{A}^{\rm{RBH}}\right|}{\mathrm{A}^{\rm{RBH}}}100\%,
\end{equation}
where $\text{RBH} = (\text{ABG}, \text{BD})$. The smaller the function $\mathrm{A}_{\rm{dif}}$ is, the more similar are the regular and singular BH areas. In Table~\ref{Tab.2}, we present the corresponding BH areas, as well as $\mathrm{A}_{\rm{dif}}$, for several pairs $(\alpha^{\rm{ABG}},\alpha^{\rm{RN}})$ for which the corresponding GCS of null geodesics are the same. As we can see, the function $\mathrm{A}_{\rm{dif}}$ increases with the increase of $\alpha$. We also note that for an ABG RBH with $\alpha \lesssim 0.5$, we have that $\mathrm{A}_{\rm{dif}} \lesssim 3\%$. Therefore, pairs $(\alpha^{\rm{ABG}},\alpha^{\rm{RN}})$ that satisfy $\sigma_{\rm{gcs}}^{\rm{ABG}} = \sigma_{\rm{gcs}}^{\rm{RN}}$ correspond to very similar BH areas, as long as we consider only low-to-moderate values of $\alpha$. This implies that we can find situations where the ACSs of ABG and RN BHs are also very similar at the low- and high-frequency regimes.
\begin{table}[hbtp!]
\centering
\begin{tabular}{||c|c|c|c|c||} 
 \hline
 $\alpha^{\rm{ABG}}$ & $\alpha^{\rm{RN}}$ & $\mathrm{A}^{\rm{ABG}}/M^{2}$ & $\mathrm{A}^{\rm{RN}}/M^{2}$ & $\mathrm{A}_{\rm{dif}}$ \\ [0.5ex] 
 \hline\hline
 $0.1$ & $0.0896$ & $50.0125$ & $50.0631$ & $0.1011\% $ \\ \hline
 $0.2$ & $0.1794$ & $49.2497$ & $49.4528$ & $0.4123\% $ \\ \hline
 $0.3$ & $0.2693$ & $47.9653$ & $48.4251$ & $0.9585\%$ \\ \hline
 $0.4$ & $0.3594$ & $46.1372$ & $46.9624$ & $1.7885\%$ \\ \hline
 $0.5$ & $0.4498$ & $43.7289$ & $45.0364$ & $2.990\%$ \\ \hline
\hline
\end{tabular}
\caption{BH areas and $\mathrm{A}_{\rm{dif}}$ for several pairs $(\alpha^{\rm{ABG}},\alpha^{\rm{RN}})$ for which the GCS of null geodesics is the same.}
\label{Tab.2}
\end{table}

As shown in Table~\ref{Tab.3}, for the BD RBH, pairs $(\alpha^{\rm{BD}},\alpha^{\rm{RN}})$ that satisfy $\sigma_{\rm{gcs}}^{\rm{BD}} = \sigma_{\rm{gcs}}^{\rm{RN}}$ correspond to very different BH areas already for moderate values of $\alpha$, i.e., $\alpha \geq 0.5$. Therefore, the differences are more pronounced in the case of BD-RN than in the case of ABG-RN. Consequently, for the BD RBH geometry, we cannot find situations in which similar GCSs lead to similar BH areas for moderate values of $\alpha$. This result contrasts with the absorption results for massless scalar fields in the ABG RBH, and help us to better understand the scalar absorption results presented in Ref.~\cite{MC2014}. Notice also that, for $\alpha < 0.5$, the location of the event horizon of the charged BHs becomes very similar to that of a Schwarzschild BH, such that it would be very unlikely to distinguish between them. For example, for $\alpha^{\rm{BD}} = 0.5$, the percentual deviation of the event horizon location compared to the Schwarzschild BH is already less than $6\%$. Therefore, the region of interest must correspond to values of $\alpha$ where $\alpha > 0.5$.
\begin{table}[hbtp!]
\centering
\begin{tabular}{||c|c|c|c|c||} 
 \hline
 $\alpha^{\rm{BD}}$ & $\alpha^{\rm{RN}}$ & $\mathrm{A}^{\rm{BD}}/M^{2}$ & $\mathrm{A}^{\rm{RN}}/M^{2}$ & $\mathrm{A}_{\rm{dif}}$ \\ [0.5ex] 
 \hline\hline
 $0.1$ & $0.077$ & $50.0417$ & $50.1163$ & $0.149\% $ \\ \hline
 $0.2$ & $0.1541$ & $49.3665$ & $49.6664$ & $0.6074\% $ \\ \hline
 $0.3$ & $0.2316$ & $48.2268$ & $48.9076$ & $1.4116\%$ \\ \hline
 $0.4$ & $0.3096$ & $46.5989$ & $47.8256$ & $2.6326\%$ \\ \hline
 $0.5$ & $0.3883$ & $44.4432$ & $46.3977$ & $4.3977\%$ \\ \hline
\hline
\end{tabular}
\caption{BH areas and $\mathrm{A}_{\rm{dif}}$ for several pairs $(\alpha^{\rm{BD}},\alpha^{\rm{RN}})$ for which the GCS of null geodesics is the same.}
\label{Tab.3}
\end{table}

Therefore, for massless test scalar fields, the key aspect to obtain similar ACSs is to find configurations in which the low- and high-frequency limits are very similar. In other words, situations where the pairs of BH charges associated with the same GCSs lead to very similar BH areas. For the scattering results, we can formulate a similar argument, but based on the similarity between the impact parameter of backscattered rays and the corresponding scattering flux intensities. For ABG and RN BHs, these quantities can be very similar for arbitrary scattering angle values, considering low-to-moderately charged BHs. However, for BD and RN BHs, even though the interference widths are very similar in the whole scattering angle domain, the scattering flux intensities are different, mainly for intermediate-to-high scattering angles.

For massive scalar fields, the situation is different. Concerning the absorption results, in the very low-frequency regime, the total ACS is essentially governed by Eq.~\eqref{lf22}, which diverges as $\omega M \rightarrow \mu M$. As this divergence dominates the behavior of the total ACS in this regime, the differences between the BH areas for the pairs in which the GCSs are the same are substantially mitigated. Therefore, for massive scalar waves, we expect to find situations in which the total ACSs of BD and RN BHs can be very similar for arbitrary frequency values and moderate values of $\alpha$, in contrast to the massless scalar wave scenario. For the differential SCS, we can formulate a heuristic argument based on the similarity between BH pairs with the same $b_{g}$ and the corresponding scattering flux intensities.

We have shown that the interference fringe width is inversely proportional to the argument of the Bessel function $(\sim 1/\omega vb_{g})$. Hence, for both massless and massive scalar waves, choosing BH configurations in which $b_{g}$ is the same leads to similar oscillatory patterns for the differential SCS. However, the scattering flux intensity is proportional to $b_{g}^{2}|db/d\theta|$. Consequently, BH pairs in which $b_{g}$ is the same do not necessarily result in BH configurations in which the corresponding flux intensity is also similar. Moreover, in Fig.~\ref{TSCSABGBDRN}, we show that although the effect of the BH charge on the scattering flux intensity is very different for each BH solution, the effect of the field mass is essentially to increase the scattering flux intensity. Therefore, we conjecture that the field mass can provide a mechanism to make the differential SCSs of BD and RN BHs similar for arbitrary scattering angle values, which is not possible when one considers massless scalar fields.

\subsection{Similar absorption and scattering properties}\label{subsec:sap}

So far, we have discussed situations in which the absorption and scattering results for singular and regular BH are similar based on analytical arguments. In what follows, we present a graphical discussion. As we pointed out in the previous section, to find situations in which the results for the ACSs are similar, we consider the pairs of normalized charges for which the critical impact parameter associated with the timelike geodesics are the same [see Eq.~\eqref{CIP}]. In the case of the differential SCS, we proceed similarly, but considering the impact parameter of backscattered rays $b_{g}$ instead of $b_{c}$. We notice that the values allowed for the equality between $b_{c}$ or $b_{g}$ depends on the value of $\omega M$, and we have chosen $\omega M > \mu M$, since we are dealing with unbounded modes. Moreover, for $\mu = 0$, we obtain the results for the massless counterpart~\cite{MC2014,PLC2020}.
\begin{figure*}[!htbp]
  \centering
  \subfigure{\includegraphics[width=0.9\columnwidth]{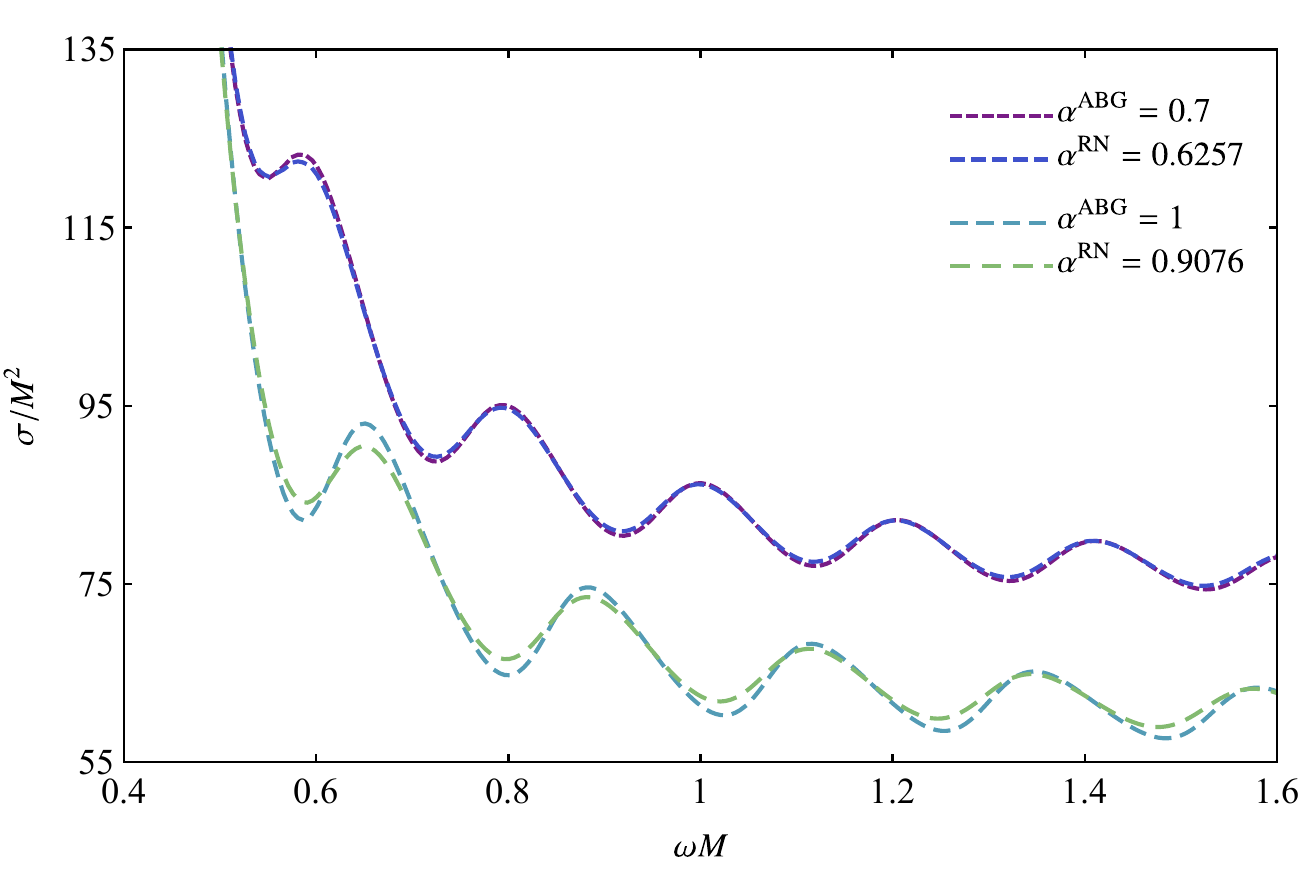}\label{c}}
    \subfigure{\includegraphics[width=0.9\columnwidth]{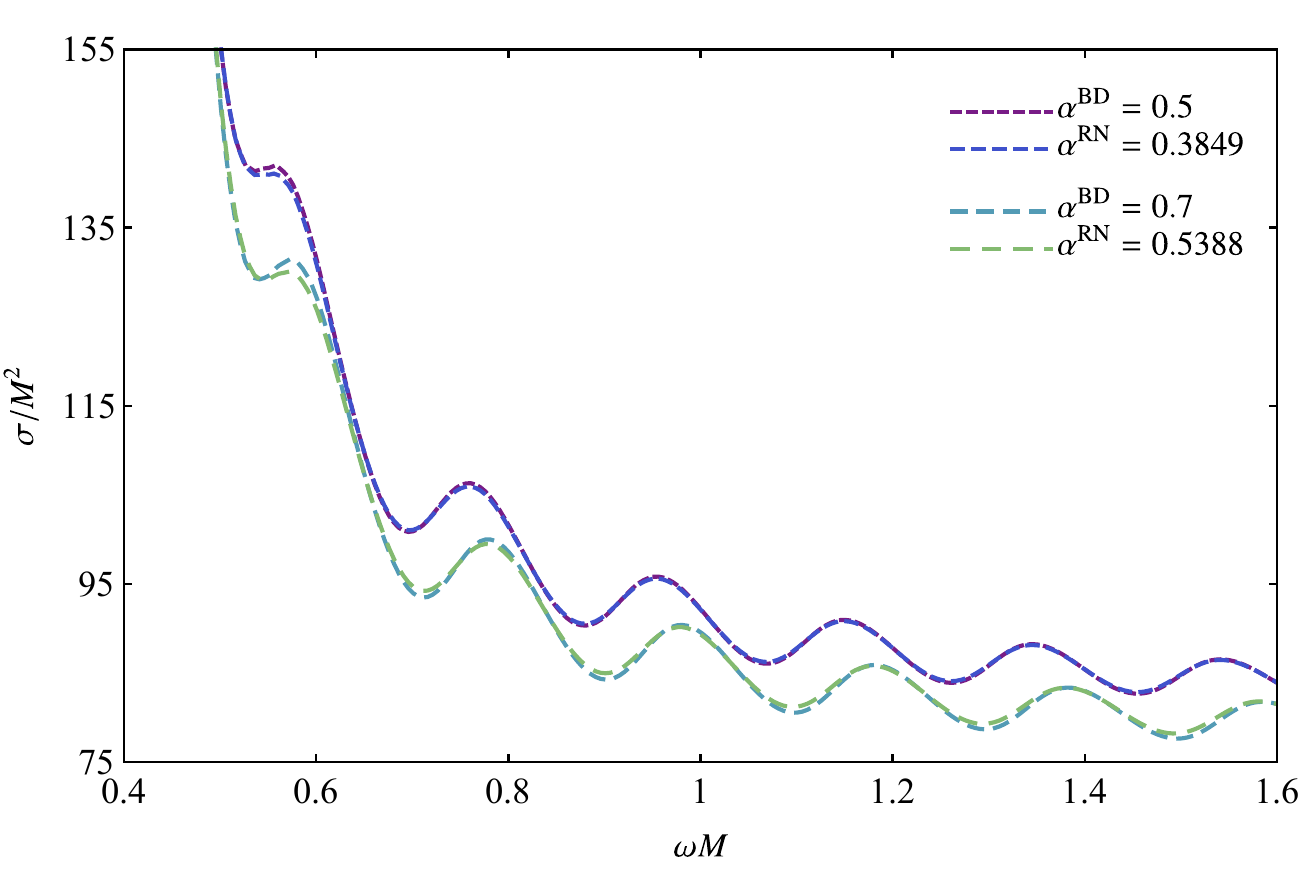}\label{d}}
     \subfigure{\includegraphics[width=0.9\columnwidth]{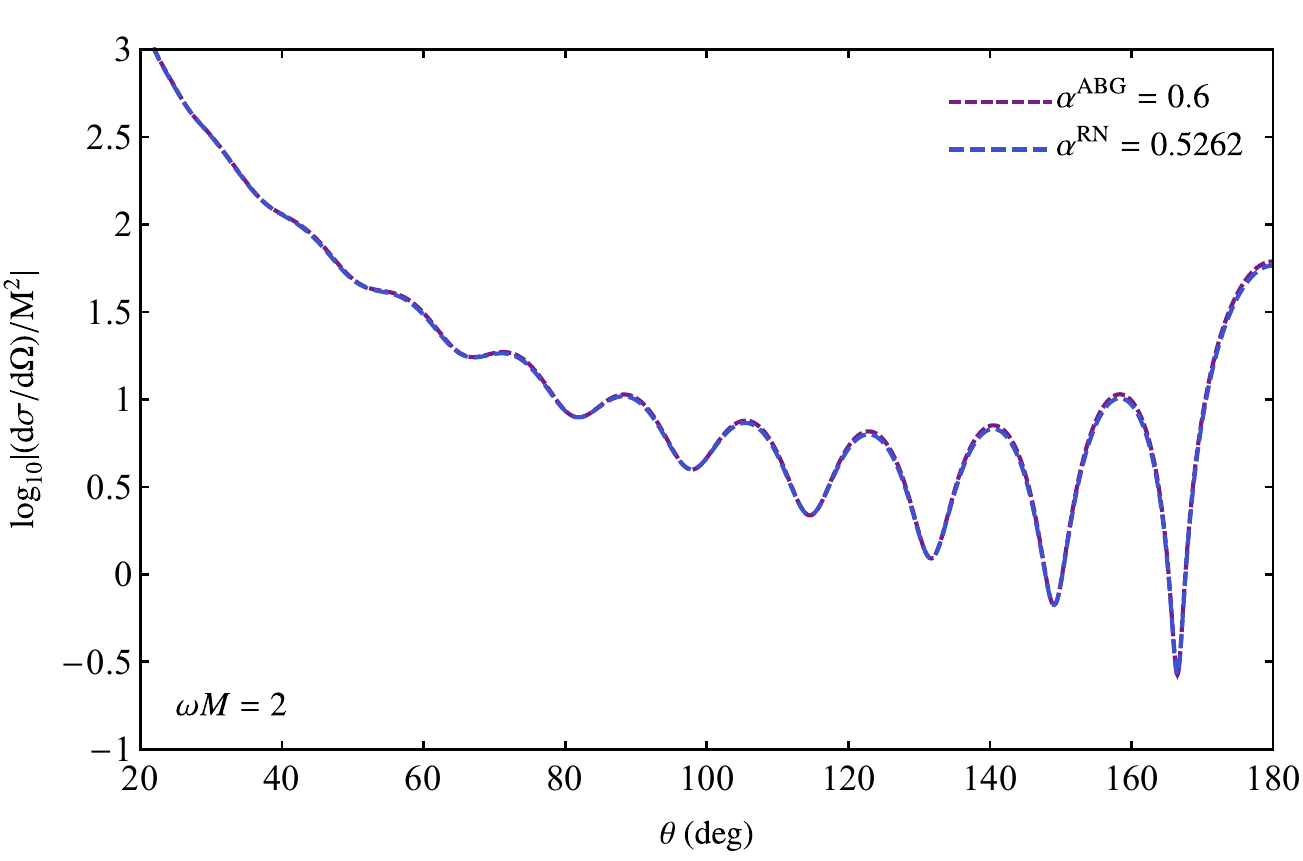}\label{e}}
  \subfigure{\includegraphics[width=0.9\columnwidth]{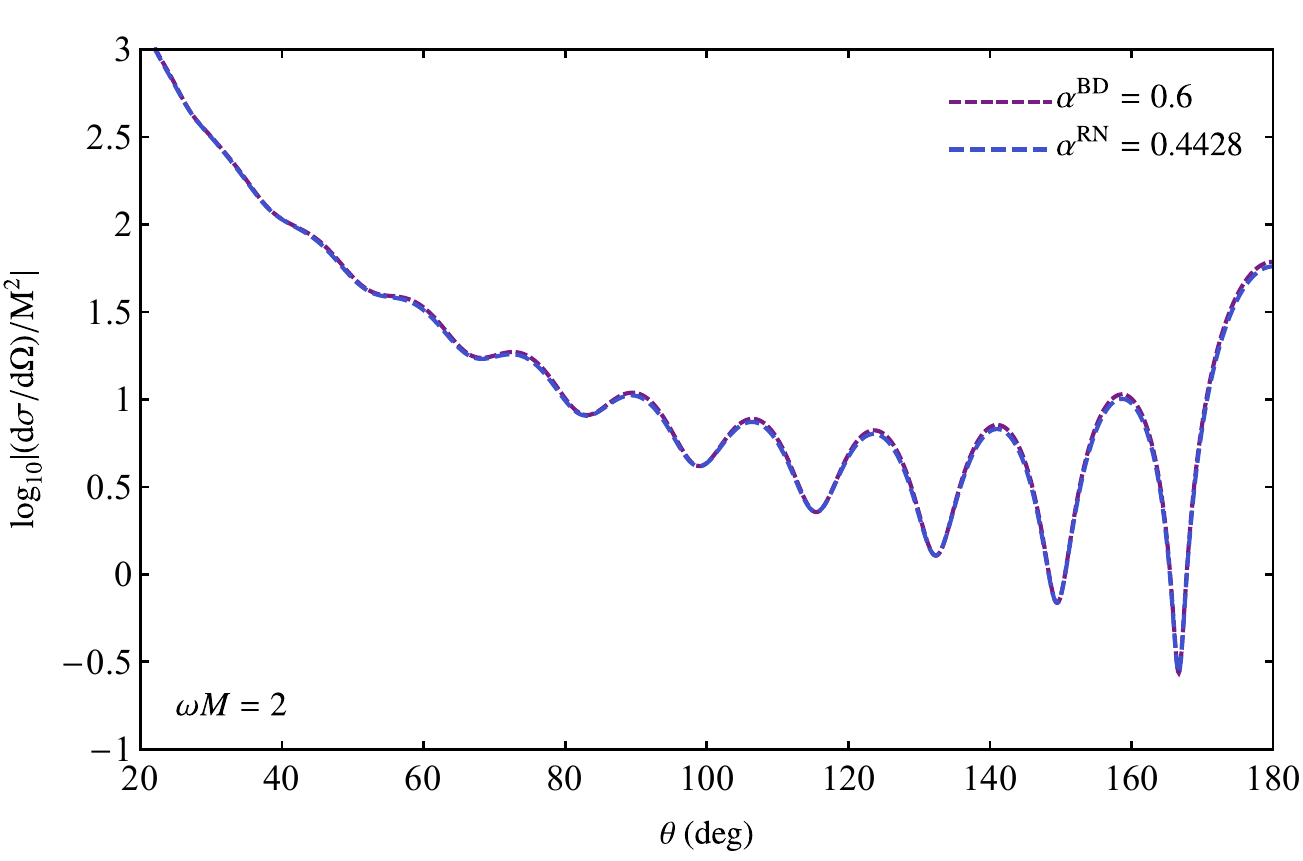}\label{f}}
  \subfigure{\includegraphics[width=0.9\columnwidth]{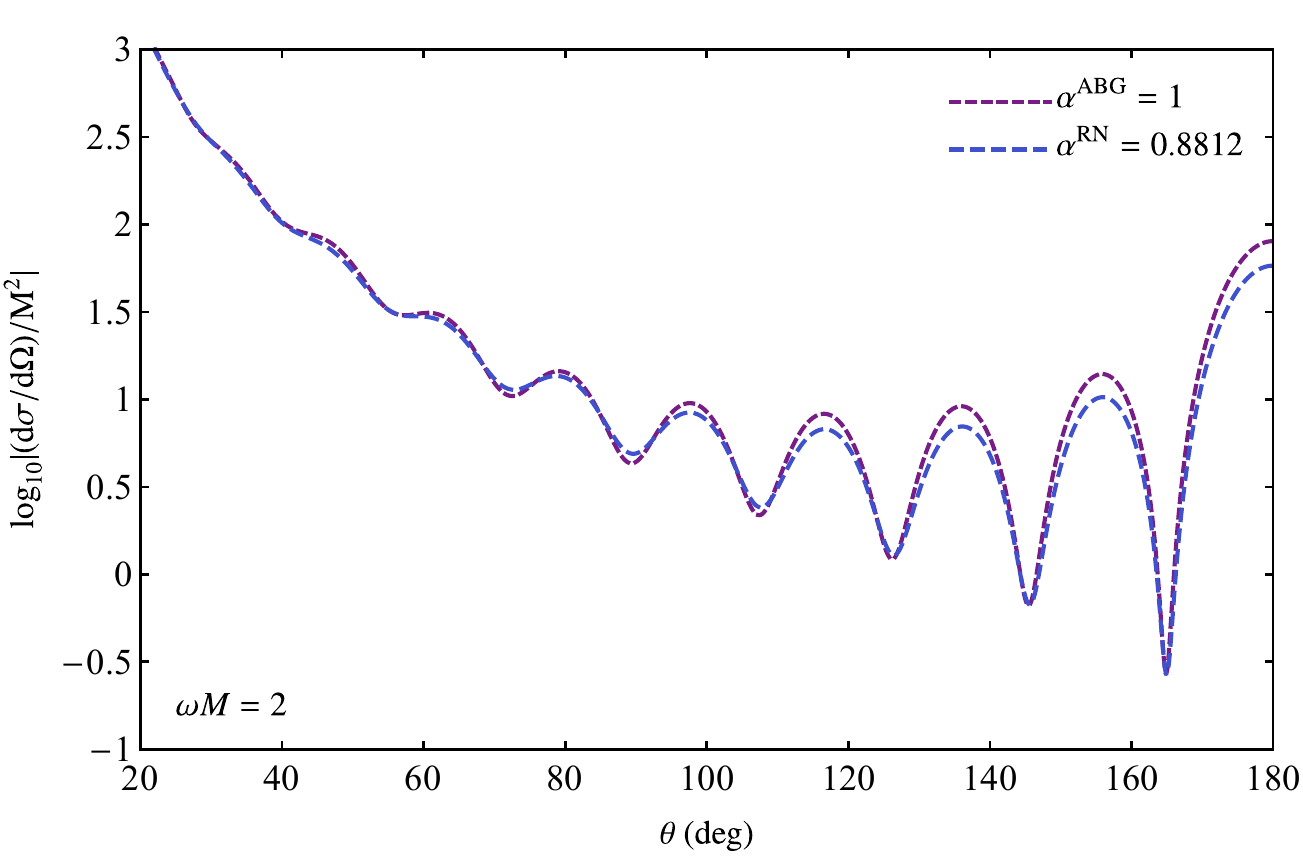}\label{g}}
    \subfigure{\includegraphics[width=0.9\columnwidth]{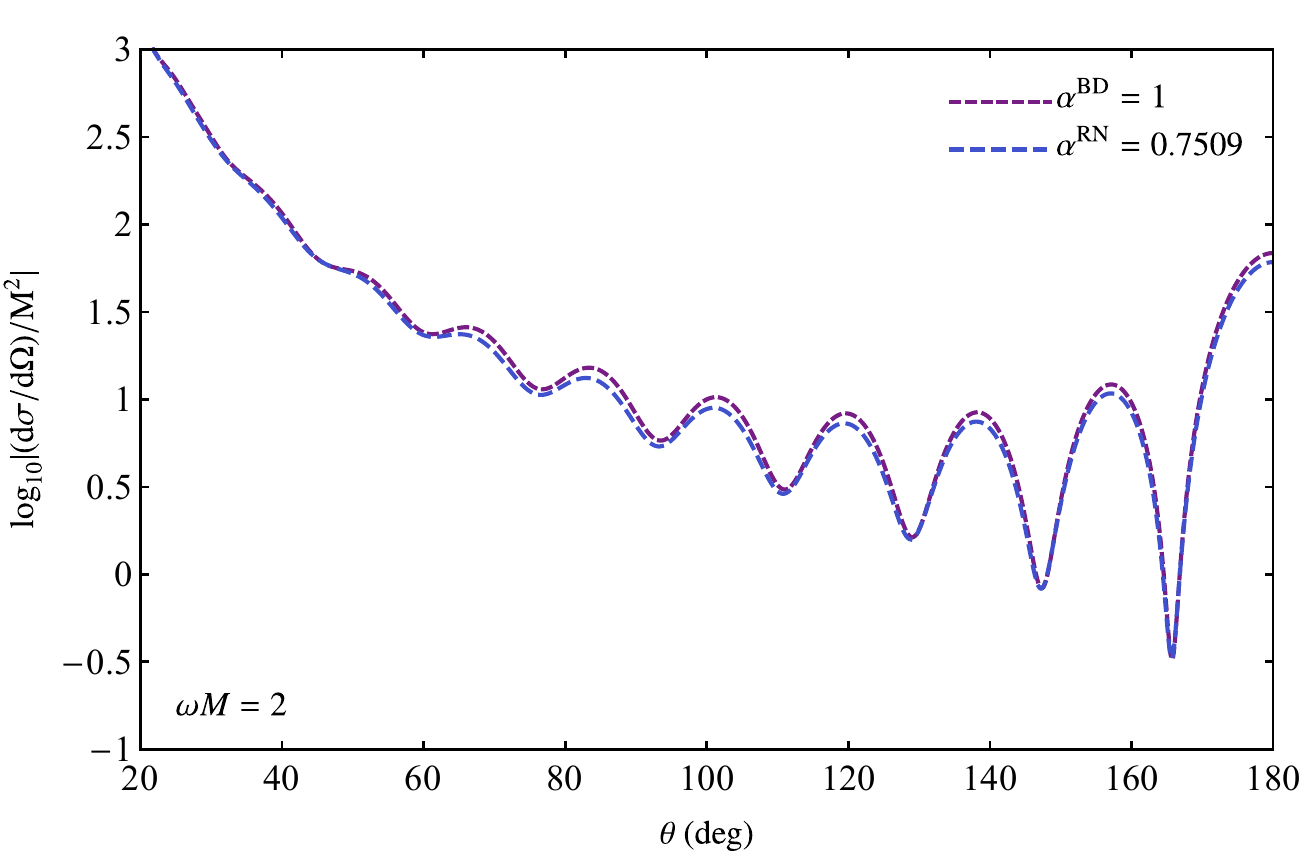}\label{h}}
\caption{Total ACS (top panels) and differential SCS (middle and bottom panels) for some pairs of $(\alpha^{\rm{ABG}}, \alpha^{\rm{RN}})$ and $(\alpha^{\rm{BD}}, \alpha^{\rm{RN}})$. In the left panels, we consider the results for the pair $(\alpha^{\rm{ABG}}, \alpha^{\rm{RN}})$, while the results for the pair $(\alpha^{\rm{BD}}, \alpha^{\rm{RN}})$ are presented on the right panels. We have chosen only pairs for which their GCS of timelike geodesics, for the absorption case, and the impact parameter of backscattered rays, for the scattering case, are equal, and considered $\mu M = 0.4$ in all scenarios.}
\label{har}
\end{figure*}

In Fig.~\ref{har}, we exhibit the absorption and differential scattering cross sections of some pairs of $(\alpha^{\rm{ABG}}, \alpha^{\rm{RN}})$ and $(\alpha^{\rm{BD}}, \alpha^{\rm{RN}})$, for which the absorption or scattering properties are similar. Concerning the absorption properties, we see that significant differences between the corresponding ACSs appear only for moderate-to-high values of the normalized charges. Remarkably, the mass of the field allows configurations for which the ACSs of BD and RN BHs are very similar in the whole frequency limit, for low- to moderately high values of the charge, in contrast to the massless case, for which the similarity can be obtained only in the high-frequency regime~\cite{MC2014}. This is in agreement with what we have conjectured in the previous section. The mass term introduces a divergence in the low-frequency regime that mitigates the difference between the corresponding BH areas, making feasible such configurations for the BD-RN case for moderate values of $\alpha$.

We notice that in the scattering scenario, the mass of the field allows new features in both ABG and BD cases. In the ABG case, it is possible to find configurations for which the total SCSs of ABG and RN BHs are very similar for low-to near-extreme values of the normalized charge. This result contrasts with the massless case, for which the similarity for arbitrary values of the scattering angle is obtained only for low-to-moderate values of the normalized charge. For its turn, in the BD case, it is possible to find situations for which the differential SCSs of BD and RN are very similar in the whole scattering angle domain, considering low-to-moderate values of the normalized charge. In the massless case, the differential SCSs of the BD RBHs can resemble only the interference widths of the RN BHs~\cite{MOC2015} --- the intensity of the scattering fluxes is different. As for the absorption results, the results obtained for the differential SCS are in agreement with what we have discussed in the previous section. The mass term affects the scattering intensity flux, allowing us to find situations where scattering properties of regular and singular BHs are similar for arbitrary scattering angle values.

\section{Concluding remarks}\label{sec:remarks}

In recent years, several studies have addressed the absorption and differential scattering cross sections of RBHs in the literature. However, most of these works have been devoted to massless test scalar fields. Here, we have presented a detailed analysis of the absorption and scattering properties of massive test scalar fields by ABG and BD RBHs, complementing the analysis of the massless counterpart investigated in Refs.~\cite{MC2014,MOC2015,PLC2020,PLC2022}. To draw comparisons with singular BHs, we have also presented our results for the RN case.

Concerning the absorption scenario, our main results for the RBHs have shown that the mass of the field typically increases the total ACS, considering fixed BH charges. For field masses satisfying $\mu M > \mu_{c} M$, the unbounded modes are strongly absorbed [see Sec.~\ref{sec:sf}, in particular,~\ref{subsec:msw}], contributing to the increase of the total ACS. As for the RN spacetime, we have also observed that the low-frequency regime of the ACS is not entirely determined by the fundamental mode $l = 0$, and the total ACS diverges in this limit, as $\omega \rightarrow \mu$. Besides that, by comparing the RBHs absorption results with those obtained for the RN case, we have shown that regular and singular BHs can have very similar absorption properties for low to nearly extreme BH charges. Moreover, our numerical results are consistent with the low- and high-frequency approximations for the ACS in the appropriate limits and with the potential barrier analyses.

Turning our attention to the scattering spectrum of RBHs, our numerical results are consistent with the classical and semiclassical approximations in their corresponding limits. For example, we observe that as $\omega \rightarrow \mu$, the divergence of the differential SCS obtained numerically is, in a sense, enhanced, in accordance with Eq.~\eqref{CSCSweakABG}. The widths of the interference fringes increase as we consider higher values of field mass for fields with velocity satisfying $v > v_{c}$, or as we increase the BH charge. The critical velocity, $v_{c}$, is obtained from the glory formula~\eqref{glory}, and we have restricted our analysis to $v > v_{c}$. We also found that for sufficiently slow massive scalar fields, the Lyapunov exponent does not increase monotonically with the increase of $\alpha$. This situation is absent for RBHs when one considers massless scalar waves. Hence, it constitutes a distinct signature of massive scalar fields. Furthermore, comparisons with the results obtained for the RN metric show that RBHs can mimic the scattering properties of singular BHs for arbitrary values of the scattering angle, considering low- to moderately high-BH charges. In the extremely charged scenario for RBHs, the intensity of the scattering flux is different, although the fringe width pattern is similar.

An interesting contribution of this work to the BH mimickers literature~\cite{MFS2021,CKO2024} is that it is easier to find configurations in which the absorption and scattering patterns of regular and singular BHs coincide when considering massive scalar fields in the vicinity of RBHs than in the massless counterpart scenario. This can be taken as an indication that, in more realistic scenarios, the absorption and scattering spectra of singular and nonsingular BHs may, in fact, not be distinguishable from the perspective of the test scalar fields. In particular, we discussed in detail the standard scenarios to find similar absorption and scattering cross sections. For the absorption results, we considered BH pairs for which the GCSs coincide, while for the scattering results, we considered scenarios where the impact parameter of the backscattered rays are the same.

For completeness, we also compared the absorption and scattering results for the ABG, BD, and RN BHs. We have shown that, for fixed values of $\alpha$ and $\mu M$, the total ACSs generally satisfy $\sigma^{\rm{BD}} > \sigma^{\rm{ABG}} > \sigma^{\rm{RN}}$. This result is consistent with the behavior of the potential barrier (see, e.g., Fig.~\ref{peffabgbdrn}). Concerning the scattering scenario, we have found that the width of the interference fringes of the RN BH is usually wider than that of the ABG RBH, which, in turn, is wider than that of the BD RBH. The increase of scalar field mass typically increases the width of the interference fringes and the scattering intensity fluxes, but the pattern observed when we increase the charge-to-mass ratio of the BHs is different.

Given the extensive literature concerning the absorption and scattering properties of compact objects, in particular considering massless scalar fields and BHs, and the similarities among some findings, we now focus on comparing our main findings with some other results reported in the literature. For Horndeski BH solutions, as we consider higher values of $\mu M$, the total ACS increases, the width of the interference fringes get wider, and the magnitude of the scattering intensity flux increases~\cite{LWJ20252}. Conversely, for the Bonanno-Reuter (BR) RBH, the graybody factors decrease as the field mass increases, implying that the total ACS decreases (fewer waves are absorbed by the BH~\cite{ZM2026}). Therefore, our results are qualitatively  more similar to Horndeski BHs than to BR RBHs. In particular, the results for the BR RBHs reinforces that the role triggered by the field mass for the absorption and scattering results is not universal. It strongly depends on the BH model.

This work deals mainly with the absorption and scattering of massive scalar fields, filling a gap in the literature. However, investigating the absorption and scattering cross sections of RBH geometries considering charged scalar fields is also a relevant task. In this case, for frequencies satisfying $\omega < q\phi(r_{+})$, where $q$ is the field charge and $\phi(r_{+})$ the radial electrostatic potential evaluated at the event horizon radius, we may observe the superradiance phenomenon in the absorption spectrum (the total ACS is negative for a certain frequency range)~\cite{BC2016,MP2024b}. Notably, the analyses presented in Refs.~\cite{MP2024b,PLC2025} also reveal that we can have unbounded superradiance, i.e., the total ACS is negative and unbounded from below, in the background of RBHs for test scalar fields, without a cavity (or mirrorlike boundary condition) to confine the superradiant modes. We can also have superradiant instability~\cite{MAP2024c,ZXZ2024,SH2025}. Furthermore, the superradiance of magnetically charged BHs in the presence of scalar fields can be relevant in the quest for magnetic monopoles in the Universe~\cite{DP2024}. Moreover, the differential SCS of singular and regular BHs for charged scalar waves has not yet been covered in the literature. We intend to address this problem in future work.

We conclude by discussing an additional physical relevance of our results. We recall that scalar fields can model dark matter candidates (see Ref.~\cite{JB2011} and references therein) and the speed of the massive scalar wave is defined by Eq.~\eqref{ADP} with $m \rightarrow \mu$ and $E \rightarrow \omega$. As the de Broglie and Compton wavelengths can be written as
\begin{equation}
\lambda_{B} = \dfrac{2\pi}{\kappa} \quad \text{and} \quad \lambda_{C} = \dfrac{2\pi}{\mu},
\end{equation}
respectively, with $\kappa = \sqrt{\omega^{2}-\mu^{2}}$, one can show that~\cite{CB2014}
\begin{equation}
\label{v2}v = \dfrac{1}{\sqrt{1+\lambda_{B}^{2}/\lambda_{C}^{2}}}.
\end{equation} 
We can consider the scenario related to the model proposed by Hui \textit{et al.} in Ref.~\cite{LH2017}, where an alternative to cold dark matter is proposed. The authors considered an extremely light scalar field with mass $m \approx 10^{-22}eV/c^{2}$ and de Broglie wavelengh $\lambda \approx 1 \text{ kpc}$. For this case, the corresponding Compton wavelength is given by $\lambda_{C} = h/\mu c \approx 0.4 \text{ pc}$, and we find that $v = 10^{-4}$. Therefore, the absorption and scattering results for (much) slower massive scalar waves can be relevant within this context. Notice, however, that faster massive scalar waves are unlikely to be used for describing dark matter candidates~\cite{RA2017}. Therefore, to explore scenarios where the absorption and scattering results are potentially relevant within this context, we can rewrite Eq.~\eqref{v2} as
\begin{equation}
\dfrac{\lambda_{B}}{\lambda_{C}} = \dfrac{\sqrt{1-v^{2}}}{v}.
\end{equation}
As we can see, for fast massive scalar waves ($v \rightarrow 1$), $\lambda_{B}$ tends to be smaller than $\lambda_{C}$. This is consistent with the case of ultrarelativistic scalar bosons, for which the kinetic energy is significantly larger than the corresponding rest mass.

Note added. \textit{As the final version of this paper was being completed, the preprint \cite{HH2026} was published on arXiv, covering some of our results for the RN case, with different choices of parameters.}

\begin{acknowledgments}

M. P. thanks Luiz Leite and Sam Dolan for useful comments and discussions. The authors would like to acknowledge Funda\c{c}\~ao Amaz\^onia de Amparo a Estudos e Pesquisas (FAPESPA),  Conselho Nacional de Desenvolvimento Cient\'ifico e Tecnol\'ogico (CNPq)  and Coordena\c{c}\~ao de Aperfei\c{c}oamento de Pessoal de N\'ivel Superior (CAPES) -- Finance Code 001, from Brazil, for partial financial support.  This research has further been supported by the European Horizon Europe staff exchange (SE) programme HORIZON-MSCA-2021-SE-01 Grant No. NewFunFiCO-101086251 and by the L’ORÉAL-UNESCO-ABC Para Mulheres na Ciência program. M. A. A. de Paula is supported by Grant No. CNPQ/PDJ 150589/2025-5.

\end{acknowledgments}

\section*{Data availability}

There are no publicly available research data or software supporting this manuscript. Requests for further information or data should be sent to the authors.

\end{document}